\documentclass[11pt]{article}
\usepackage{graphicx, caption, subcaption, braket, amsmath, amsfonts} 
\usepackage[dvipsnames]{xcolor}
\usepackage{graphicx}
\usepackage{float}

\usepackage{tikz}
\usetikzlibrary {intersections}
\usetikzlibrary{decorations.pathreplacing}
\usetikzlibrary{decorations.pathmorphing,decorations.markings,decorations.pathmorphing,decorations.markings,decorations.pathmorphing,decorations.markings}

\newcommand{\red}[1]{\textcolor{red}{#1}}

\usepackage{jheppub}

\preprint{LCTP-24-22}

\title{Quantum-Corrected Hawking Radiation from Near-Extremal Kerr-Newman Black Holes}

\usepackage{float}

\author[a]{Sabyasachi Maulik}
\author[b,c]{Xin Meng}
\author[c,d]{Leopoldo A. Pando Zayas }

\emailAdd{mauliks@iitk.ac.in,xinmeng@umich.edu, lpandoz@umich.edu}

\affiliation[a]{Department of Physics, Indian Institute of Technology Kanpur, Kalyanpur, Kanpur, Uttar Pradesh 208016, India}

\affiliation[b]{University of Science and Technology of China, Hefei, Anhui 230026, China}

\affiliation[c]{Leinweber Center for Theoretical Physics, 
University of Michigan, Ann Arbor, MI 48109, USA}

\affiliation[d]{The Abdus Salam International Centre for Theoretical Physics, 34014 Trieste, Italy}

\abstract{Near-extremal black holes have a long AdS$_2$ throat in their near-horizon region. Quantum fluctuations  in the throat region are effectively governed by a quantum version of Jackiw-Teitelboim gravity  with matter and are strongly coupled at low temperatures.  We investigate how these quantum fluctuations affect the spectrum of emission of particles during Hawking radiation. We systematically consider the cases of Kerr and Kerr-Newman black holes for emission of scalar particles and discuss photon and graviton emission from the Kerr background. We find that at very low temperatures the quantum fluctuations radically change the nature of particle emission. Unlike the generic suppression of particle emission in the Reissner-Nordstr\"om case, we uncover that for particles with non-vanishing angular momentum, the quantum-corrected emission can be substantially enhanced with respect to the standard semiclassical result.}

\date{}

\begin{document}

\maketitle
\section{Introduction}

Black holes provide ideal laboratories for the study of quantum aspects of gravity. Classically, black holes are completely black and fully characterized by their mass, charge and angular momentum. Hawking radiation is arguably the first result demonstrating that quantum-mechanically black holes behave very differently from the classical general-relativistic expectation. Hawking radiation has led to a number of important conceptual puzzles, pitting general relativity and quantum mechanics against each other. Hawking's computation is intrinsically semiclassical, it considers quantum fields in a fixed classical gravitational background \cite{Hawking:1975vcx}. As a first important approximation, this approach is quite justified but it is natural to wonder about the effects of including quantum fluctuations of the metric. In this manuscript we take some steps in this direction.

There are by now various technical approaches to black hole Hawking radiation, we will focus on the spontaneous emission approach pioneered by Press and Teukolsky more than half a century ago \cite{Press:1972zz}. This approach is appropriately suited to include the effects of quantum fluctuations for the class of black holes that we address in this manuscript. Computations of particle emission from black holes have an old and venerated history, dating back to foundational works in the 1970's  \cite{Press:1972zz,Press:1973zz,Teukolsky:1974yv,Teukolsky:1973ha,Teukolsky:1972my,Page:1976df,Page:1976ki}. Most of these results have remained intact for over four decades. Recently, however, an interesting connection with other problems in mathematical physics has provided the ability to perform certain black hole computations exactly in the frequency of the emitted particles \cite{Aminov:2020yma,Bonelli:2021uvf}, we will also incorporate these recent developments into our computations.

The main conceptually new ingredient that we bring to the question of Hawking radiation is the fact that the quantum fluctuations in the near-horizon region of near-extremal black holes are now well understood. Near-extremal black holes develop a long AdS$_2$ throat in their near-horizon region. There has recently been impressive progress in the understanding of the quantum nature of the AdS$_2$ region 
\cite{Almheiri:2014cka, Jensen:2016pah, Maldacena:2016upp}. The progress was originally achieved in the simple framework of Jackiw-Teitelboim (JT) gravity \cite{Jackiw:1984je, Teitelboim:1983ux}. This understanding of the nature of the quantum fluctuations was later applied to higher-dimensional near-extremal black holes and led to important modifications to the low-temperature thermodynamics of black holes \cite{Iliesiu:2020qvm, Heydeman:2020hhw, Boruch:2022tno, Iliesiu:2022onk}.

More recently, the physics of the throat region has been an essential ingredient in determining  logarithmic-in-temperature corrections to the thermodynamics of rotating black holes in asymptotically flat spacetimes \cite{Kapec:2023ruw, Rakic:2023vhv}. These results were shown to be universal in \cite{Maulik:2024dwq} by extending the one-loop quantum corrections to the thermodynamics for asymptotically AdS spacetimes and black holes in various dimensions. The corrections to the thermodynamics are obtained by regulating a set of zero modes following recent suggestions of turning on a small but finite temperature in the geometry \cite{Iliesiu:2022onk, Banerjee:2023quv, Banerjee:2023gll}. Other related explorations motivated by the quantum nature of fluctuations near the throat region have been reported recently \cite{Bai:2023hpd, Daguerre:2023cyx,Kapec:2024zdj, Kolanowski:2024zrq,Liu:2024gxr,Heydeman:2024fgk,Heydeman:2024ezi,Liu:2024qnh}.

Beyond understanding the effects of quantum fluctuations in the thermodynamics, it is only natural to track their influence in more dynamical processes, such as Hawking radiation.  Indeed, recently the effect of these strong quantum fluctuations have been taken into account for charged, spherically symmetric black holes \cite{Brown:2024ajk}. Our main goal in this manuscript is to extend the framework of quantum-corrected Hawking radiation, introduced in \cite{Brown:2024ajk}, to the context of rotating black holes. 

One natural motivation to consider rotating black holes is that astrophysical near-extremal rotating black holes, that is,  black holes rotating at near-maximal rate, have been established in the range from solar mass (Cygnus X-1, \cite{Zhao_2021}, one of the component stars of the binary X-ray system Cygnus X-1 is a black hole) to billion solar masses (M87${}^*$) \cite{10.1093/mnrasl/slz176}.

The remainder of the manuscript is organized as follows. In section \ref{Sec:Prelims} we review the literature devoted to the computations of the spectra of particle emission by black holes; we also include some recent results from the literature in mathematical physics that allow for an exact computation of the greybody factors. Section \ref{Sec:Quantum} introduces the paradigm of quantum corrections arising by incorporating the influence of strong quantum fluctuations in the throat. In section \ref{Sec:Scalar}, we present the results for scalar particles, while in section \ref{Sec:PhotonGraviton} we briefly address aspects of photon and graviton emission. We summarize our results and conclude in Section \ref{Sec:Conclusions}. We relegate some technical material to a series of appendices. 

\section{Black Hole Preliminaries}\label{Sec:Prelims}

\subsection{Kerr-Newman geometry}

The metric and gauge field of a Kerr-Newman black hole with mass $M$, angular momentum $J = M a$, and electric charge $Q$ is written in Boyer-Lindquist coordinates as
\begin{align}
    ds^2 &= -\frac{\Delta}{\rho^2}\left(dt - a \sin^2\theta d\phi\right)^2 + \frac{\rho^2}{\Delta} dr^2 + \rho^2 d\theta^2 + \frac{1}{\rho^2} \sin^2\theta \left(a dt - \left(r^2 + a^2\right) d\phi \right)^2, \label{eq:KN_metric} \\
    A &= -\frac{Q r}{\rho^2}\left(dt - \sin^2\theta d\phi\right)
\end{align}
where
\begin{align}
\begin{split}
    \Delta &= \left(r^2 + a^2\right) - 2 M r + Q^2, \qquad 
    \rho^2 = r^2 + a^2\cos^2\theta.
\end{split}
\end{align}
The black hole has two event horizons, which can be obtained as the two real, positive roots of $\Delta = 0$; they are given by
\begin{equation}
    r_{\pm} = M \pm \sqrt{M^2 - a^2 - Q^2}.
\end{equation}
The Hawking temperature, classical entropy, electric potential, and angular velocity of the horizon are
\begin{subequations}
\begin{align}
        T_{H} &= \frac{r_{+} - r_{-}}{4\pi\left(r_{+}^2 + a^2\right)},\qquad S = \pi\left(r_{+}^2 + a^2\right), \nonumber \\   \Phi &= \frac{Q r_{+}}{r_{+}^2 + a^2},\qquad    \Omega_{H} = \frac{a}{r_{+}^2 + a^2}.
\end{align}
\end{subequations}
For future reference, we also define a dimensionless Hawking temperature
\begin{equation}
    \tau_{H} \equiv \frac{r_{+} - r_{-}}{r_{+}}.
\end{equation}

\subsubsection{Low-temperature expansion}

In the extremal limit, the Hawking temperature $\left(T_{H}\right)$ of the black hole slowly approaches zero, and the two event horizons $r_{\pm}$ merge into each other. An extremal Kerr-Newman black hole has the lowest possible mass in the solution space, which is given by \red{$\left. M\right |_{T_{H} = 0} = \sqrt{a^2 + Q^2}$}.

By expanding the thermodynamic quantities around their extremal values, we may find the effective temperature scale below which semiclassical description is expected to break down \cite{Preskill:1991tb,Maldacena:1998uz,Page:2000dk}. The near-extremal expansion depends on the choice of ensemble; we mention here the results obtained in the canonical ensemble (where the electric charge $Q$ and angular momentum $J$ are held fixed), and the grand canonical ensemble (where the electric potential $\Phi$, and the angular velocity are held constant).\\

\noindent (i) \emph{Near-extremal expansion in the canonical ensemble:} 

In the canonical ensemble, we hold the electric charge $Q$ and the angular momentum $J$ of the black hole fixed, and expand all other parameters around their extremal values. At small but non-vanishing temperature the event horizon radii admit the following expansion
\begin{subequations}
    \begin{align}
        r_{+} &= r_{0} + 2\pi \left(r_{0}^2 + a_{0}^2 \right) T_{H} + O\left(T_{H}^2\right),\\
        r_{-} &= r_{0} - 2\pi \left(r_{0}^2 + a_{0}^2 \right) T_{H} + O\left(T_{H}^2\right),
    \end{align}
\end{subequations}
which automatically leads to the near-extremal expansion of the energy or mass of the black hole
\begin{equation}
    M = r_{0} + 2\pi^2 r_{0} \left(r_{0}^2 + a_{0}^2 \right) T_{H}^2,
\end{equation}
and the thermodynamic potentials
\begin{subequations}
\begin{align}
    \Phi &= \frac{r_{0}^2 \sqrt{r_{0}^2-a_{0}^2}}{a_{0}^2+r_{0}^2} + \frac{4 \pi  a_{0}^2 r_{0} \sqrt{r_{0}^2-a_{0}^2}}{a_{0}^2+r_{0}^2} T_{H},\\
    \Omega_{H} &= \frac{a_{0}}{a_{0}^2+r_{0}^2}-\frac{4 \pi  a_{0} r_{0}}{a_{0}^2+r_{0}^2} T_{H}.
\end{align}
\end{subequations}
In these expansions, $a_{0}$ and $r_{0}$ represent the extremal values of the angular momentum parameter $a$ and the outer event horizon radius $r_{+}$. They are related to the constant quantities $Q$ and $J$ by
\begin{align}
    r_{0}^2 = \frac{1}{2} \left(Q^2 + \sqrt{Q^4 + 4 J^2} \right),\quad a_{0}^2 = \frac{1}{2} \left(-Q^2 + \sqrt{Q^4 + 4 J^2} \right).
\end{align}
Finally, the Bekenstein-Hawking entropy of the Kerr-Newman black hole can be expanded near extremality as
\begin{equation} \label{eq:entropy_lowT_canonical}
    S = \pi \left(r_{0}^2 + a_{0}^2 \right) + 4\pi^2 r_{0} \left(r_{0}^2 + a_{0}^2 \right) T_{H}.
\end{equation}
Therefore, the energy $E_{\text{brk}}$ at which semiclassical thermodynamics is supposed to breaks down or equivalently the Schwarzian modes become strongly coupled is given by \cite{Sachdev:2019bjn, Kolanowski:2024zrq}
\begin{equation}
    E_{\text{brk}}^{-1} = \left| \frac{1}{4\pi^2} \frac{\partial S}{\partial T_{H}} \right|_{T_{H} = 0} = r_{0} \left(r_{0}^2 + a_{0}^2 \right).
\end{equation}
For a Kerr black hole $\left( Q = 0 \right)$, this yields $E^{\text{Kerr}}_{\text{brk}} = \frac{1}{2 J^{\frac{3}{2}}}$. While for a Reissner-Nordstr\"{o}m black hole $\left(J = 0 \right)$, we get $E^{\text{RN}}_{\text{brk}} = \frac{1}{Q^{3}}$; commensurate with the observations in \cite{Kapec:2023ruw, Rakic:2023vhv, Iliesiu:2020qvm}.\\

\noindent (ii) \emph{Near-extremal expansion in the grand canonical ensemble:}

Similar low temperature expansions can be obtained by holding the thermodynamic potential $\Phi$ and $\Omega$ constant, while letting the conserved charges $Q$ and $J$ depend on temperature. In this case, we find
\begin{subequations}
    \begin{align}
        r_{+} &= r_{0} + \frac{2 \pi  r_{0}^2 \left(r_{0}^4 - a_{0}^4 \right)}{2 a_{0}^4 - 5 a_{0}^2 r_0^2 + r_{0}^4} T_{H} + O\left(T_{H}^2\right),\\
        r_{-} &= r_{0} - \frac{2 \pi \left(a_{0}^2+r_{0}^2\right) \left(4 a_{0}^4-9 a_{0}^2 r_{0}^2+r_{0}^4\right)}{2 a_{0}^4-5 a_{0}^2 r_{0}^2+r_{0}^4}  T_{H} + O\left(T_{H}^2\right).
    \end{align}
\end{subequations}
We also note the low temperature expansion of the angular momentum per unit mass, $a = \frac{J}{M}$
\begin{equation}
    a = a_{0} + \frac{4 \pi r_{0}^3 a_{0} \left(a_{0}^2+r_{0}^2\right)}{2 a_{0}^4-5 a_{0}^2 r_{0}^2+r_{0}^4} T_{H} + O\left(T_{H}^2\right).
\end{equation}

This time the two extremal parameters $r_{0}$ and $a_{0}$ are determined by their relation with the thermodynamic potentials
\begin{equation}
    \Phi = \frac{r_{0}^2 \sqrt{r_{0}^2-a_{0}^2}}{a_{0}^2+r_{0}^2}, \quad \Omega = \frac{a_{0}}{a_{0}^2 + r_{0}^2}.
\end{equation}
The other expressions are somewhat longer, and to avoid clutter we only mention here the expansions of the black hole mass, electric charge and entropy
\begin{align}
    M &= r_{0}-\frac{4 \pi \left(a_{0}^6 - a_{0}^4 r_{0}^2 - 2  a_{0}^2 r_{0}^4 \right)}{2 a_{0}^4-5 a_{0}^2 r_{0}^2+r_{0}^4} T_{H}^2,\\
    Q &= \sqrt{r_{0}^2-a_{0}^2} - \frac{4 \left(\pi  a_{0}^6 r_{0}-\pi  a_{0}^2 r_{0}^5\right)}{\sqrt{r_{0}^2-a_{0}^2} \left(2 a_{0}^4-5 a_{0}^2 r_{0}^2+r_{0}^4\right)} T_{H},\\
    S &= \pi  \left(a_{0}^2+r_{0}^2\right) + \frac{4 \pi ^2 r_{0}^3 \left(a_{0}^2+r_{0}^2\right)^2}{2 a_{0}^4-5 a_{0}^2 r_{0}^2+r_{0}^4} T_{H}.
\end{align}
Hence, the breakdown scale is given by
\begin{equation}
    E_{\text{brk}}^{-1} = \left| \frac{r_{0}^3 \left(a_{0}^2+r_{0}^2\right)^2}{2 a_{0}^4-5 a_{0}^2 r_{0}^2+r_{0}^4} \right|.
\end{equation}
In the grand canonical ensemble for a Kerr black hole, $E_{\text{brk}}^{\text{Kerr}} = 4 \Omega_{H}^3$, and for a Reissner-Nordstr\"{o}m black hole, $E_{\text{brk}}^{\text{RN}} = \frac{1}{\Phi^{3}}$. As already anticipated in \cite{Preskill:1991tb,Maldacena:1998uz,Page:2000dk}, the physics near the scale $E_{\text{brk}}$ receives quantum corrections whose implications for Hawking radiation is the main goal of this manuscript.\\

\noindent \emph{Estimate of $E_{\text{brk}}$ for astrophysical black holes:} 
In April 2019, The Event Horizon Telescope (EHT) \cite{2024A&A...681A..79E} reported the first ever horizon-scale images of a black hole, resolving the central compact radio source in the giant elliptical galaxy M 87. The black hole M87* has a mass around 5.37 billion times the solar mass M$_{\odot}$ $= 1.9884 \times 10^{30}$ kg \cite{2023ApJ...945L..35L}. In Planck units, the mass of M87* is estimated to be
\begin{equation*}
    M_{87^{*}} = 5.37 \times 10^{9}\times \frac{1.9884 \times 10^{30}}{2.1764 \times 10^{-8}} = 4.9061 \times 10^{47}.
\end{equation*}
The dimensionless rotation parameter $a_{*} = \frac{a}{M}$ for M87* has been estimated to be: $a_{*} = 0.90$ \cite{10.1093/mnrasl/slz176}. Therefore, the black hole angular momentum is $J = Ma = 2.1663 \times 10^{95}$.

Using these parameter values, we may estimate that for M87* the breakdown scale for a with fixed $J$ is 
\begin{equation*}
E_{\text{brk}} = \frac{1}{2 J^{\frac{3}{2}}} = 4.95909 \times 10^{-144}.
\end{equation*}
in Planck units. This is much smaller than any realistic temperature in the universe, suggesting that semiclassical thermodynamics is extremely robust for an astrophysical black hole like M87*. Similarly, the breakdown scale with fixed $\Omega_{H}$ is $E_{\mathrm{brk}} = \frac{1}{4 \Omega_{H}^3} = 1.0426 \times 10^{-144}$.

Another well-studied example of a black hole in astrophysics is the Cygnus-X1, itself a member of a binary with mass around 21.2 M$_{\odot}$ \cite{Orosz:2011ApJ, Miller-Jones:2021AA}, and dimensionless spin parameter $a_{*} = 0.99$ \cite{Zhao_2021}. For such a black hole, the breakdown scales with fixed $J$ and fixed $\Omega_{H}$ are $6.888 \times 10^{-119}$ and $4.493\times 10^{-119}$, respectively.

\subsection{Perturbation of Kerr black hole: The Teukolsky Master Equation} \label{sec:Teukolsky_master_equations}

In this section we revisit the perturbation equations which govern the dynamics of fields of spin projections $s = 0, \pm \frac{1}{2}, \pm 1, \pm 2$ moving in a black hole background.  This is relatively straightforward for a static and spherically symmetric background metric (e.g. Schwarzschild and Reissner-Nordstr\"{o}m black hole), where the field equations can be easily decoupled in terms of a radial mode $\left(R_{\ell}\right)$ and the spherical harmonics $\left(Y_{\ell m}\right)$.  For backgrounds which include rotation such as Kerr or Kerr-Newman, the spacetime metric is more complicated. Moreover, the replacement of spherical symmetry by axial symmetry means that a separation into spherical harmonics is no longer possible. Fortunately, Teukolsky \cite{Teukolsky:1972my, Teukolsky:1973ha} showed that if one works directly in terms of curvature invariants, the perturbation equations for general spin decouple and separate for a Kerr spacetime. For the more general Kerr-Newman metric, the Klein-Gordon equation $\left(s = 0\right)$ and Dirac equation $\left(s = \frac{1}{2}\right)$ are separable \cite{Page:1976jj}, but difficulty remains over the decoupling of electromagnetic and gravitational perturbations \cite{Chandrasekhar:1983bk}. Hence, we set the electric charge of the black hole $Q = 0$, and discuss the master perturbation equation derived by Teukolsky for fields of general spin on a Kerr black hole. For details and further discussion on this topic, the reader is referred to \cite{Breuer:1975bk, Chandrasekhar:1983bk, Brito:2015oca}.

Teukolsky's approach is based on the Newman-Penrose (NP) formalism \cite{Newman:1961qr}. In this formalism, one introduces a tetrad of null vectors $\mathbf{l, n, m,}$ and $\mathbf{m^{*}}$ at each point in spacetime. All tensors are projected onto this null tetrad. The vectors $\mathbf{l}$ and $\mathbf{n}$ are real, and $\mathbf{m}$ and $\mathbf{m^{*}}$ are complex conjugate of each other. The NP tetrad must satisfy the orthogonality relations
\begin{equation*}
    \mathbf{l.n = 1},\quad \mathbf{m.m^{*}=-1},\quad \text{all other dot products zero}.
\end{equation*}
For the Kerr metric in Boyer-Lindquist coordinates \eqref{eq:KN_metric} (with Q = 0), the NP tetrad may be conveniently chosen as \cite{Teukolsky:1973ha}
\begin{align}
\begin{split}
    l^{\mu} &= \left[\frac{r^2 + a^2}{\Delta}, 1, 0, \frac{a}{\Delta} \right],\quad n^{\mu} = \frac{1}{2\rho^2}\left[r^2 + a^2, -\Delta, 0, a \right],\\ m^{\mu} &= \frac{1}{\sqrt{2}\left(r + i a \cos\theta\right)}\left[i a \sin\theta, 0, 1, i\csc\theta \right].
\end{split}
\end{align}
Using these expressions, Teukolsky derived a single master equation valid equally well for a spin $s$ field $\left(s = 0, \pm \frac{1}{2}, \pm 1, \pm 2\right)$ $\psi \left(t, r, \theta, \phi\right)$ in the Kerr background
\begin{equation} \label{eq:Teukolsky_master}
    \begin{split}
        &\left[\frac{\left(r^2 + a^2 \right)^2}{\Delta} - a^2\sin^2\theta \right] \frac{\partial^2\psi}{\partial t^2} + \frac{4 M a r}{\Delta} \frac{\partial^2\psi}{\partial t \partial \phi} + \left[\frac{a^2}{\Delta} - \frac{1}{\sin^2\theta} \right] \frac{\partial^2\psi}{\partial\phi^2}\\ &- \Delta^{-s}\frac{\partial}{\partial r}\left(\Delta^{s+1} \frac{\partial\psi}{\partial r}\right) - \frac{1}{\sin\theta}\frac{\partial}{\partial\theta}\left(\sin\theta\frac{\partial\psi}{\partial\theta}\right) - 2s\left[\frac{a\left(r-M\right)}{\Delta} + \frac{i\cos\theta}{\sin^2\theta} \right] \frac{\partial\psi}{\partial\phi}\\ &-2s\left[\frac{M\left(r^2-a^2\right)}{\Delta} - r - i a \cos\theta \right]\frac{\partial\psi}{\partial t} + \left(s^2\cot^2\theta - s \right)\psi = 0.
    \end{split}
\end{equation}
By further using a separation of variables Ansatz for $\psi\left(t, r, \theta, \phi\right)$ in terms of a radial mode and the spin-weighted \emph{spheroidal} harmonics
\begin{equation}
    \psi\left(t, r, \theta, \phi\right) = e^{-i \omega t + i m \phi} S_{\ell m}\left(\theta\right) R_{s}\left(r\right),
\end{equation}
the master equation can be separated into two ODEs
\begin{align}
    \Delta \frac{d^2R_{s}}{dr^2} + \left(s+1\right) \frac{d\Delta}{dr} \frac{dR_{s}}{dr} + \left(\frac{K^2 - 2 i s \left(r - M\right) K}{\Delta} + 2 a m \omega - K_{\ell} +4 i s \omega r\right) R_{s} &= 0, \label{eq:Teukolsky_radial-1} \\
    \partial_{z} \left(1-z^2\right)\partial_{z}S_{\ell m} + \left(K_{\ell} + s - a^2\omega^2\left(1-z^2\right) - \frac{\left(m+s z\right)^2}{1-z^2} - 2 a \omega s z \right) S_{\ell m} &= 0. \label{eq:Teukolsky_angular-1}
\end{align}
In these equations $K = \left(r^2 + a^2\right)\omega - a m$, and $z = \cos\theta$. Solutions to the angular equation \eqref{eq:Teukolsky_angular-1} are known in the literature as spin-weighted spheroidal harmonics. For $a\omega = 0$, the spin-weighted spheroidal harmonics reduce to spin-weighted spherical harmonics $Y_{\ell m}\left(\theta, \phi\right)$. In this case the spheroidal eigenvalues or the angular separation constants $K_{\ell}$ are known analytically: $K_{\ell} \left(a = 0\right) = \ell\left(\ell + 1\right) - s\left(s + 1\right)$. For general rotation parameter, determining this constant is a nontrivial problem.

Finally, let us mention how the Teukolsky master wavefunction $\psi$ is related to the perturbation fields. In NP formalism, the perturbation variables are usually some curvature tensor contracted on the tetrad legs; e.g. the electromagnetic perturbations are described by
\begin{equation*}
    \mathbf{\Phi}_{0} = F_{\mu\nu} l^{\mu} m^{\nu},\quad \mathbf{\Phi}_{2} = F_{\mu\nu} m^{* \mu} n^{\nu}.
\end{equation*}
The gravitational quantities of interest are some contracted components of the Weyl tensor
\begin{equation*}
    \mathbf{\Psi}_{0} = - C_{\mu\nu\lambda\sigma} l^{\mu} m^{\nu} l^{\lambda} m^{\sigma},\quad \mathbf{\Psi}_{4} = - C_{\mu\nu\lambda\sigma} n^{\mu} m^{*\nu} n^{\lambda} m^{*\sigma},\quad
\end{equation*}
The complete relations for each value of $s$ is given in table \ref{table:Teukolsky_relations}. 
\begin{table}[h!]
    \centering
    \begin{tabular}{|c|c|c|c|c|}
        \hline
        $s$ & 0 & $\left(\frac{1}{2}, -\frac{1}{2}\right)$ & $\left(1, -1 \right)$ & $\left(2, -2 \right)$ \\
        $\psi$ & $\Phi$ & $\left(\chi_{0}, \varrho^{-1}\chi_{1} \right)$ & $\left(\Phi_{0}, \varrho^{-2} \Phi_{2} \right)$ & $\left(\Psi_{0}, \varrho^{-4} \Psi_{4} \right)$ \\
        \hline
    \end{tabular}
    \caption{Teukolsky wavefunction $\psi\left(t, r, \theta, \phi \right)$ for each value of spin $s$. The spin coefficient $\varrho = -\frac{1}{r - i a \cos\theta}$. $\chi_{0}$ and $\chi_{1}$ refer to the components of the fermion wavefunction along dyad legs.}
    \label{table:Teukolsky_relations}
\end{table}

\subsection{From Teukolsky master equation to confluent Heun equation}

In order to compute the emission rate of neutral particles from a Kerr black hole, one should divide the spacetime outside the outer horizon into a near and far region, solve the equation \eqref{eq:Teukolsky_radial-1} in the two regions, and calculate the greybody factor using appropriate matching conditions between the regions, before normalizing with the correct probability or energy flux. This is easier said than done because for an arbitrary frequency $\left(\omega\right)$ of the emitted particle, the division of the spacetime into near- and far-horizon regions is often unclear, even when the black hole is near extremality. In fact, most of the previous works \cite{Maldacena:1997ih, Bredberg:2009pv, Hartman:2009nz} assume low temperature as well as low frequency to deduce some analytically tractable results.
 
 In this regard, the recent work \cite{Bonelli:2021uvf} made a remarkable contribution by mapping the radial and angular Teukolsky equations to the confluent Heun equation (CHE). Further, the confluent Heun equation is also shown to match the BPZ equation of the two-dimensional chiral conformal field theory \cite{Belavin:1984vu} in the Nekrasov-Shathashvili (NS) double-scaling limit \cite{Nekrasov:2009rc}. By using results from conformal field theory to connect the asymptotic expansion of CHE near two regular singular points, the authors of \cite{Bonelli:2021uvf} managed to provide exact expressions for the greybody factor of an uncharged, rotating black hole. For low frequency and small rotation parameter, their result matches with the universal answer of \cite{Maldacena:1997ih}. 
 
 Below we briefly describe the mapping from Teukolsky equation to confluent Heun equation. In the next sections, we use the connection coefficients derived from CHE to calculate the classical greybody factor as well as the quantum emission rates of neutral particles from a rotating black hole. A complete analysis of the connection formulae using the relation between CHE and BPZ equations is beyond the scope of the present paper, and for that the reader is invited to consult \cite{Bonelli:2021uvf} and references therein. We are going to operate as users of those results and collect a few computational details in  appendix \ref{appendix:CHE}.\\

For the radial Teukolsky equation \eqref{eq:Teukolsky_radial-1}, let us define
\begin{equation}
    x = \frac{r - r_{-}}{r_{+} - r_{-}},\quad \chi(x) = \Delta^{\frac{s+1}{2}} (r) R_{s} (r).
\end{equation}
Due to this change of variables the inner and outer horizon radii are situated at $x = 0$, and $x = 1$, respectively, and $r \to \infty$ corresponds to $x \to \infty$. We obtain the differential equation
\begin{equation} \label{eq:Heun_radial}
    \frac{d^{2}\chi(x)}{dx^2} + V_{r}(x) \chi(x) = 0,\quad \text{with potential } V_{r}(x) = \frac{1}{x^2\left(x-1\right)^2} \sum_{i=0}^{4} \hat{A}_{i}^{r} x^{i}.
\end{equation}
The coefficients $\hat{A}_{i}^{r}$ depend on the black hole parameters $\left(M, J\right)$ and the frequency, spin and angular momentum of the perturbation. Their explicit expressions are collected below \cite{Bonelli:2021uvf}
\begin{equation}
    \begin{split}
        \hat{A}_{0}^{r} &= \frac{a^2 \left(1 - m^2\right) - M^2 + 4 a m M \omega r_{-} + 4 M^2 \omega^2 \left(a^2 - 2 M^2\right) + 8 M^3 \left(M^2 - a^2\right) \omega^2}{4\left(a^2-M^2\right)}\\ &\hspace{1em} + \left(i s\right) \frac{a m \sqrt{M^2 - a^2} - 2 a^2 M \omega + 2 M^2 \omega r_{-}}{2 \left(a^2 - M^2\right)} - \frac{s^2}{4},\\
        \hat{A}_{1}^{r} &= \frac{4\left(a^2-M^2\right)\left(K_{\ell} - a^2\omega^2\right) + \left(8amM\omega + 16 a^2 M \omega^2 - 32 M^3 \omega^2 \right) \sqrt{M^2 - a^2} + 4 a^4 \omega^2}{4\left(a^2 - M^2\right)}\\ &\hspace{1em} + \frac{8 M^4 \omega^2 - 9 a^2 M^2 \omega^2}{a^2 - M^2} + \left(i s \right) \left(-i + \frac{\left(2 a^2 \omega - a m\right) \sqrt{M^2 - a^2}}{a^2 - M^2} \right) + s^2,\\
        \hat{A}_{2}^{r} &= -K_{\ell} - 4 a^2\omega^2 + 12 M^2\omega^2 - 12 M\omega^2\sqrt{M^2-a^2} + \left(i s\right)\left(i - 6\omega\sqrt{M^2 - a^2}\right) - s^2,\\
        \hat{A}_{3}^{r} &= 8 \left(a^2-M^2\right)\omega^2 + \left(8 M\omega^2 + 4 i s \omega\right)\sqrt{M^2-a^2},\\
        \hat{A}_{4}^{r} &= 4\left(M^2-a^2\right)\omega^2.
    \end{split}
\end{equation}

For the angular equation \eqref{eq:Teukolsky_angular-1}, one may instead define
\begin{equation}
    x = \frac{1+z}{2},\quad y(x) = \frac{\sqrt{1-z^2}}{2} S_{\ell m}.
\end{equation}
Since originally $z = \cos\theta$, the above change maps $\theta = 0$ to $x = 1$, and $\theta = \pi$ to $x = 0$. The angular equation then also takes the form of a Schr\"{o}dinger equation
\begin{equation} \label{eq:Heun_angular}
    \frac{d^2y(x)}{dx^2} + V_{\text{ang}}(x) y(x) = 0,\quad \text{where } V_{\text{ang}}(x) = \frac{1}{x^2\left(x-1\right)^2} \sum_{i=0}^{4} \hat{A}_{i}^{\theta} x^{i},
\end{equation}
with the coefficients $\hat{A}_{i}^{\theta}$ of the angular part being given by
\begin{equation}
    \begin{split}
        \hat{A}_{0}^{\theta} &= -\frac{1}{4}\left(-1+m-s\right)\left(1+m-s\right),\\
        \hat{A}_{1}^{\theta} &= K_{\ell} + s + 2a\omega s - ms + s^2,\\
        \hat{A}_{2}^{\theta} &= -K_{\ell} + a^2\omega^2 - s - \left(a\omega + s\right)\left(5a\omega + s\right),\\
        \hat{A}_{3}^{\theta} &= 4 a \omega \left(2a\omega + s\right),\\
        \hat{A}_{4}^{\theta} &= -4a^2\omega^2.
    \end{split}
\end{equation}
When written as Schr\"{o}dinger equations, both the radial and angular equations \eqref{eq:Heun_radial} and \eqref{eq:Heun_angular} share the same singularity structure. They have two regular singularities at $x = 0, 1$, and an irregular singular point at $x = \infty$. Such an equation in Mathematics literature is known as the confluent Heun equation \cite{Ronveaux:1995hn}.

The same confluent Heun equation arises as a limit to the BPZ equation of two dimensional chiral conformal field theory
\begin{equation}
    \frac{d^2\psi}{dx^2} + \frac{1}{x^2 \left(x-1\right)^2} \sum_{i=0}^{4} A_{i} x^{i} = 0,
\end{equation}
with the coefficients
\begin{equation}
    \begin{split}
        A_{0} &= \frac{1}{4} - a_{1}^2,\\
        A_{1} &= -\frac{1}{4} + E + a_{1}^2 - a_{2}^2 - m_{3}\Lambda,\\
        A_{2} &= \frac{1}{4} - E + 2 m_{3} \Lambda - \frac{\Lambda^2}{4},\\
        A_{3} &= -m_{3}\Lambda + \frac{\Lambda^2}{2},\\
        A_{4} &= -\frac{\Lambda^2}{4}.
    \end{split}
\end{equation}
The mapping is completed by giving a dictionary to connect the parameters of the CFT with the black hole parameters. There are two different dictionaries for the radial and angular perturbation equations. For the radial equation \eqref{eq:Teukolsky_radial} or equivalently \eqref{eq:Heun_radial} the dictionary is given by

\begin{equation}\label{eq:Heun_radial_dictionary}
    \begin{split}
        E &= \frac{1}{4} + K_{\ell} + s\left(s+1\right) - 8M^2\omega^2 - \left(2M\omega^2 + i s \omega\right)\left(r_{+}-r_{-}\right),\\
        a_{1} &= -i\frac{\omega - m\Omega_{H}}{4\pi T_{H}} + 2iM\omega + \frac{s}{2}, \qquad 
        a_{2} = -i\frac{\omega - m\Omega_{H}}{4\pi T_H}-\frac{s}{2}\qquad \\
        m_{3} &= -2iM\omega + s,\qquad 
        \Lambda = -2i\omega\left(r_{+}-r_{-}\right).
    \end{split}
\end{equation}
On the other hand, the dictionary for the angular part of the equation \eqref{eq:Heun_angular} is

\begin{equation}\label{eq:Heun_angular_dictionary}
    \begin{split}
        E &= \frac{1}{4} + K_{\ell} + s(s+1) -2a\omega s,\\
        a_{1} &= -\frac{m-s}{2},\qquad 
        a_{2} = -\frac{m+s}{2},\\
        m_{3} &= - s, \qquad \Lambda = 4 a \omega.
    \end{split}
\end{equation}

\subsection{Semiclassical greybody factor}

Particles emitted near the horizon of a black hole usually exhibit a thermal black body spectrum \cite{Hawking:1975vcx}. However, the spacetime outside the black hole acts as some sort of a potential barrier for the emitted particles -- hence the emission spectrum measured by an observer at infinity is no longer thermal, and is corrected as $\frac{\sigma\left(\omega\right)}{\exp\left(\frac{\omega - m \Omega_{H}}{T_{H}}\right) \pm 1}$, where $\sigma\left(\omega\right)$ is called the `greybody factor'. The greybody factor in asymptotically flat spacetime is the same as the absorption coefficient of the black hole, which is the ratio of the flux of particles absorbed by the black hole after overcoming the potential barrier to the flux of incoming radiation. There is a renowned universal result for this quantity in the low energy sector \cite{Das:1996we}.

The absorption cross-section of bosonic particles emitted with very low energy $\left(M \omega \ll 1\right)$ were computed by Starobinsky and Churilov in \cite{Starobinskil:1974nkd}. Later, semiclassical greybody factors were reported numerically for $s=0,\frac{1}{2}, 1, 2$ in \cite{Page:1976df,Page:1976ki}  for the case of uncharged and slowly rotating black hole. One of our necessary steps in this work is to bring those results to bear with the new understanding of the situation of near-extremal black holes.

\begin{figure}[H]
\begin{centering}
    \begin{tikzpicture}

		\filldraw[black] (2,5) circle[radius=50pt] node[red]{\textbf{\huge{M,J,Q}}};
        \draw [color=green, decorate, decoration={snake,amplitude=0.7mm,segment length=1mm}] (2,5) circle[radius=54pt];

		\draw[<-,thick,blue] (8,5.3)--(9,5.3) node[yshift=5pt,xshift=-5pt]{\large{Ingoing}}node[yshift=15pt,xshift=-5pt,orange]{\large{$m$,$j$,$e$}};
		\draw[<-,thick,blue] (9,4.7)--(8,4.7) node[yshift=-5pt,xshift=20pt]{\large{Outgoing}};
		\draw[<-,thick] (4,5)--(5,5) node[yshift=5pt,xshift=-9pt,blue]{\large{Absorbed}};
	\end{tikzpicture}
\caption{The black hole radiation absorption process that leads to the computation of the greybody factors via spontaneous emission. The near-horizon region is depicted in green anticipating the strong quantum fluctuations that will be incorporated in section \ref{Sec:Quantum}.}
\end{centering}
\end{figure}

\subsubsection{Scalar}

In the calculation of absorption coefficient or greybody factor, one is usually interested in a solution of the wave equation which is ingoing at the horizon $x = 1$. A solution near the horizon obeying the ingoing boundary condition can be expressed as
\begin{equation}
    R^{s=0}_{\ell m}\left(r \to r_{+}\right) \sim \left(r - r_{+}\right)^{-i\frac{\omega - m\Omega_{H}}{4\pi T_{H}}}.
\end{equation}
Therefore, the radial Teukolsky wavefunction $\chi(x) = \Delta^{\frac{1}{2}} R_{0}(x)$ near the horizon is given by
\begin{equation}
    \chi(x) \sim x^{\frac{1}{2}} \left(x-1\right)^{\frac{1}{2} - i\frac{\omega - m\Omega_{H}}{4\pi T_{H}}}.
\end{equation}
To write the last expression, we used $x - 1 = \frac{r - r_{+}}{r_{+} - r_{-}}$, and $\Delta = \left(r-r_{+}\right)\left(r-r_{-}\right) = x\left(x-1\right)\left(r_{+}-r_{-}\right)^2$. When the radial Teukolsky equation \eqref{eq:Teukolsky_radial-1} for $s=0$ is written as a Schr\"{o}dinger equation \eqref{eq:Heun_radial}, it admits a conserved ``probability flux''
\begin{equation}
    \varphi = \text{Im}\ \psi^{\dagger}\partial_{x}\psi,\quad \text{for $x$ on the real line}.
\end{equation}
With this definition we obtain
\begin{equation}
    \varphi_{\text{abs}} = -\frac{\omega - m \Omega_{H}}{4\pi T_{H}},
\end{equation}
as the flux absorbed on the horizon. Near infinity, the ingoing part of the wave can be found by using the connection formula for Heun, as described in equation (\ref{Eq:DOZZ}) in appendix \ref{appendix:CHE} 

\begin{equation} \label{eq:ingoing_psi_scalar}
    \chi(x) = M_{\alpha_{2+}, \alpha_{-}}\ \Lambda^{-\frac{1}{2} - \tilde{a} + m_{3}}\ \mathcal{A}_{\alpha_{-}, m_{3}+}\ x^{m_{3}} e^{\frac{\Lambda x}{2}} \frac{\langle \Delta_{\alpha_{-}}, \Lambda_{0}, m_{0}+ \lvert V_{\alpha_{2}}(1) \rvert \Delta_{\alpha_{1}} \rangle}{\langle \Delta_{\alpha}, \Lambda_{0}, m_{0}+ \lvert V_{\alpha_{2+}}(1) \rvert \Delta_{\alpha_{1}} \rangle} + \left(\alpha \to - \alpha \right),
\end{equation}
Here, $\Lambda = -2 i \omega \left(r_{+} - r_{-}\right)$, and other terms have been explained in in appendix \ref{appendix:CHE}. The parameter $\tilde{a}$ is defined through the Matone relation \cite{Matone:1995rx}
\begin{equation}
    E = \tilde{a}^2 - \Lambda \partial_{\Lambda} \mathcal{F}^{\text{inst}},
\end{equation}
where $\mathcal{F}^{\text{inst}}$ is the instanton part of Nekrasov-Shatashvili free energy. It is possible to determine $\mathcal{F}^{\text{inst}}$ perturbatively order by order in $\Lambda$, which in turn can be substituted in the Matone relation to obtain an approximate solution for $\tilde{a}(E)$ -- ultimately relating everything to the black hole parameter via the dictionary \eqref{eq:Heun_radial_dictionary}. The relevant quantities up to leading order in $\Lambda$ are given by (see equation (175) of \cite{Bonelli:2021uvf})
\begin{align}
    \tilde{a}(E) &= \sqrt{E} - \frac{\frac{1}{4} - E + a_{1}^2 - a_{2}^2}{\sqrt{E}\left(1 - 4E\right)} m_{3} \Lambda + O\left(\Lambda^2\right),\\
    \begin{split}
    \mathcal{F}^{\text{inst}} &= -\frac{\frac{1}{4} - \tilde{a}^2 - m_{1} m_{2}}{\sqrt{E}\left(1 - 4E \right)} m_{3} \Lambda + O\left(\Lambda^2\right),\\
    &\simeq \frac{-\frac{1}{4} + E + m_{1} m_{2}}{\sqrt{E} \left(1 - 4 E \right)} m_{3} \Lambda + O\left(\Lambda^2\right). \label{eq:Finst_leading}
    \end{split}
\end{align}
This is good enough for us because $\Lambda \sim T_{H}$, and ultimately we are interested in the physics close to extremality, {\it e.g.}, at vanishing temperature, $T_H$. Using equation \eqref{eq:ingoing_psi_scalar} the incoming flux from infinity can be computed as
\begin{equation} \label{eq:flux_from_infty}
    \begin{split}
        \varphi_{\text{in}} &= \text{Im}\frac{\Lambda}{2} \left| M_{\alpha_{2+}, \alpha_{-}}\ \mathcal{A}_{\alpha_{-}, m_{3}+}\ \frac{\langle \Delta_{\alpha_{-}}, \Lambda_{0}, m_{0}+ \lvert V_{\alpha_{2}}(1) \rvert \Delta_{\alpha_{1}} \rangle}{\langle \Delta_{\alpha}, \Lambda_{0}, m_{0}+ \lvert V_{\alpha_{2+}}(1) \rvert \Delta_{\alpha_{1}} \rangle} \Lambda^{-\frac{1}{2} - \tilde{a} + m_{3}} + \left(\alpha \to - \alpha \right) \right|^2,\\
        &= -\frac{1}{2} \left| \frac{\Gamma\left(1 + 2\tilde{a}\right) \Gamma\left(2\tilde{a}\right) \Gamma\left(1 + 2 a_{2} \right)}{\Gamma\left(\frac{1}{2} + m_{3} + \tilde{a} \right) \prod_{\pm} \Gamma\left(\frac{1}{2} \pm a_{1} + a_{2} + \tilde{a} \right)} \frac{\langle \Delta_{\alpha_{-}}, \Lambda_{0}, m_{0}+ \lvert V_{\alpha_{2}}(1) \rvert \Delta_{\alpha_{1}} \rangle}{\langle \Delta_{\alpha}, \Lambda_{0}, m_{0}+ \lvert V_{\alpha_{2+}}(1) \rvert \Delta_{\alpha_{1}} \rangle} \Lambda^{\tilde{a} + m_{3}} + \left(\tilde{a} \to -\tilde{a} \right) \right|^2,\\
        &= -\frac{1}{2}\left| \frac{\Gamma\left(1 + 2\tilde{a}\right) \Gamma\left(2\tilde{a}\right) \Gamma\left(1 + 2 a_{2} \right) \left(-2 i \omega \left(r_{+} - r_{-} \right) \right)^{\tilde{a} - 2 i M \omega} e^{-i\omega\left(r_{+}-r_{-}\right)}}{\Gamma\left(\frac{1}{2} - 2 i M \omega + \tilde{a} \right) \Gamma\left(\frac{1}{2} - i\frac{\omega - m\Omega}{2\pi T_{H}} + 2 i M \omega + \tilde{a} \right) \Gamma\left(\frac{1}{2} - 2 i M \omega + \tilde{a} \right)} + \left(\tilde{a} \to -\tilde{a}\right)\right|^2.
    \end{split}
\end{equation}
The last line is obtained by using (see equation (189) of \cite{Bonelli:2021uvf})
\begin{equation}
    \frac{\langle \Delta_{\alpha_{-}}, \Lambda_{0}, m_{0}+ \lvert V_{\alpha_{2}}(1) \rvert \Delta_{\alpha_{1}} \rangle}{\langle \Delta_{\alpha}, \Lambda_{0}, m_{0}+ \lvert V_{\alpha_{2+}}(1) \rvert \Delta_{\alpha_{1}} \rangle} = e^{-i\omega\left(r_{+} - r_{-}\right)} \left. \exp\left(\partial_{a_{1}} \mathcal{F^{\text{inst}}}\right) \right|_{a_{1} = \tilde{a}, a_{2} = -\tilde{a}},
\end{equation}
and noting that $\mathcal{F}^{\text{inst}}$ is actually independent of $a_{1, 2}$ at leading order in $\Lambda$ (equation \eqref{eq:Finst_leading}). The semiclassical greybody factor is, therefore
\begin{equation}
\begin{split}
    \sigma = \frac{\varphi_{\text{abs}}}{\varphi_{\text{in}}} = &\frac{\omega - m\Omega_{H}}{2\pi T_{H}} \times\\ &\left| \frac{\Gamma\left(1 + 2\tilde{a}\right) \Gamma\left(2\tilde{a}\right) \Gamma\left(1 + 2 a_{2} \right) \left(-2 i \omega \left(r_{+} - r_{-} \right) \right)^{-\tilde{a} - 2 i M \omega} e^{-i\omega\left(r_{+}-r_{-}\right)}}{\Gamma\left(\frac{1}{2} - 2 i M \omega + \tilde{a} \right) \Gamma\left(\frac{1}{2} - i\frac{\omega - m\Omega}{2\pi T_{H}} + 2 i M \omega + \tilde{a} \right) \Gamma\left(\frac{1}{2} - 2 i M \omega + \tilde{a} \right)} + \left(\tilde{a} \to -\tilde{a}\right)\right|^{-2}.
\end{split}
\end{equation}
To be understood as a series in $\Lambda$, or equivalently $T_{H}$.

\subsubsection{Photon and graviton}
For perturbation fields of spin $s > 1$, the semiclassical greybody factor may be found by considering energy fluxes \cite{Teukolsky:1974yv}. Let us very briefly sketch how this is done. On general grounds, one expects that an electromagnetic perturbation has the following form near the black hole
\begin{equation} \label{eq:near_photon_wavefunction}
    \psi\left(t, r, \theta, \phi \right) \sim e^{-i\omega t + i m \phi} S_{\ell m}\left(\theta\right) Y_{\text{abs}}\ \Delta^{-1} \left(r - r_{+}\right)^{-i \frac{\omega - m\Omega_{H}}{4\pi T_{H}}},
\end{equation}
which gives the energy flux absorbed by the black hole
\begin{equation}
    \frac{d^2E_{\text{abs}}}{dt d\Omega} = \frac{S^2_{\ell m}\left(\theta\right)}{2\pi} \frac{\omega}{8M r_{+} \left(\omega - m\Omega_{H}\right)} \left|Y_{\text{abs}}\right|^2.
\end{equation}

Similarly, at large $r$ the Teukolsky master wavefunction is supposed to exhibit the behaviour
\begin{equation} \label{eq:larger_photon_wavefunction}
    \psi\left(t, r, \theta, \phi \right) \sim e^{-i\omega t + i m \phi} S_{\ell m}\left(\theta\right) Y_{\text{in}}\ r^{-1-2 i M \omega}\ e^{-i \omega r},
\end{equation}
leading to the energy flux coming from infinity
\begin{equation}
    \frac{d^2E_{\text{abs}}}{dt d\Omega} = \frac{S^2_{\ell m}\left(\theta\right)}{2\pi} \frac{1}{4} \lvert Y_{\text{in}} \rvert^2.
\end{equation}
The semiclassical greybody factor is obtained by taking the ratio of the two fluxes, and using the normalization condition $\int d\Omega\ S^2_{\ell m} (\theta) = 1$
\begin{equation}
    \sigma = \frac{dE_{\text{abs}}/dt}{dE_{\text{in}}/dt} = \frac{\omega}{2 M r_{+} \left(\omega - m\Omega_{H}\right)} \left| \frac{Y_{\text{abs}}}{Y_{\text{in}}} \right|^2.
\end{equation}
The results of \cite{Bonelli:2021uvf} fit very nicely in this framework. As already said in equation \eqref{eq:ingoing_psi_scalar}, the incoming part of the radial wavefunction far from the black hole behaves as
\begin{equation}
    \chi\left(x\right) = \tilde{Y}_{in}\ \Lambda^{-\frac{1}{2} - \tilde{a} + m_{3}}\ e^{\frac{\Lambda x}{2}}\ x^{m_{3}} + \left(\alpha \to -\alpha\right),
\end{equation}
we are now using $\tilde{Y}_{\text{in}}$ to denote the coefficient coming from connection formulae of confluent Heun equation
\begin{equation}
    \tilde{Y}_{\text{in}} = M_{\alpha_{2+}, \alpha_{-}}\ \mathcal{A}_{\alpha_{-}, m_{3}+}\ \frac{\langle \Delta_{\alpha_{-}}, \Lambda_{0}, m_{0}+ \lvert V_{\alpha_{2}}(1) \rvert \Delta_{\alpha_{1}} \rangle}{\langle \Delta_{\alpha}, \Lambda_{0}, m_{0}+ \lvert V_{\alpha_{2+}}(1) \rvert \Delta_{\alpha_{1}} \rangle}.
\end{equation}
All other parameters are related to those of the black hole by the radial dictionary in equation \eqref{eq:Heun_radial_dictionary}. We use the transformation rule $\chi\left(x\right) = \Delta^{\frac{s+1}{2}} R_{s} (x)$, and $x = \frac{r - r_{-}}{r_{+} - r_{-}}$, with $s = 1$ to obtain
\begin{equation}
    R^{s=1}_{\ell m} \left(r \to \infty\right) = \tilde{Y}_{\text{in}}\,\left(r_{+} - r_{-} \right)^{-m_{3}} \Lambda^{-\frac{1}{2} - \tilde{a} + m_{3}}\ r^{-2iM\omega-1}\ e^{-i\omega r} + \left(\alpha \to -\alpha\right),
\end{equation}
with the expected asymptotic behaviour \eqref{eq:larger_photon_wavefunction}. For the wavefunction near the horizon \eqref{eq:near_photon_wavefunction}, we set $Y_{\text{abs}} = \left(r_{+} - r_{-}\right)^{i\frac{\omega - m\Omega_{H}}{4\pi T_{H}}}$ by appropriately choosing the constant of integration. Finally, the semiclassical greybody factor is given by
\begin{equation}
    \sigma = \frac{\omega}{2 M r_{+} \left(\omega - m\Omega_{H}\right)} \left|Y_{\text{in}}\right|^{-2},
\end{equation}
with
\begin{equation}
    Y_{\text{in}} = \tilde{Y}_{\text{in}}\, \left(r_{+} - r_{-} \right)^{-m_{3}}\ \Lambda^{-\frac{1}{2} - \tilde{a} + m_{3}} + \left(\alpha \to -\alpha\right),
\end{equation}

or directly expressing everything in terms of the black hole parameters
\begin{equation}
\begin{split}
    \sigma = &\frac{\omega}{2 M r_{+} \left(\omega - m\Omega_{H}\right)} \times \\ &\left| \frac{\Gamma\left(1 + 2\tilde{a}\right) \Gamma\left(2\tilde{a}\right) \Gamma\left(1 + 2 a_{2} \right) \left(-2 i \omega \right)^{\frac{1}{2} - \tilde{a} - 2iM\omega} \left(r_{+} - r_{-} \right)^{-\frac{1}{2} - \tilde{a}} e^{-i\omega\left(r_{+}-r_{-}\right)}}{\Gamma\left(\frac{1}{2} - 2 i M \omega + \tilde{a} \right) \Gamma\left(\frac{1}{2} - i\frac{\omega - m\Omega}{2\pi T_{H}} + 2 i M \omega + \tilde{a} \right) \Gamma\left(\frac{1}{2} - 2 i M \omega + \tilde{a} \right)} + \left(\tilde{a} \to -\tilde{a}\right) \right|^{-2}
\end{split}
\end{equation}
the second term in the parentheses happens to be of order $O(r_{+}-r_{-})^{\frac{1}{2}+\tilde{a}}$, this is a higher order term for near-extremal black holes since the Hawking temperature $T_{H} \sim \left(r_{+} - r_{-}\right)$. Hence for our subsequent analysis for low temperature black hole we can ignore this term.

Similarly, the greybody factor for graviton is
\begin{equation}
    \begin{split}
    \sigma = &\frac{\omega^{3}}{\left(2 M r_{+} \right)^{3}\left(\omega - m\Omega_{H}\right)\left(\left(\omega - m\Omega_{H}\right)^{2}+4\pi^{2}T_{H}^{2}\right)} \left|Y_{\text{in}}\right|^{-2}
    \\=&\frac{\omega^{3}}{\left(2 M r_{+} \right)^{3}\left(\omega - m\Omega_{H}\right)\left(\left(\omega - m\Omega_{H}\right)^{2}+4\pi^{2}T_{H}^{2}\right)}\times
    \\ &\left| \frac{\Gamma\left(1 + 2\tilde{a}\right) \Gamma\left(2\tilde{a}\right) \Gamma\left(1 + 2 a_{2} \right) \left(-2 i \omega \right)^{\frac{1}{2} - \tilde{a} - 2iM\omega} \left(r_{+} - r_{-} \right)^{-\frac{1}{2} - \tilde{a}} e^{-i\omega\left(r_{+}-r_{-}\right)}}{\Gamma\left(\frac{1}{2} - 2 i M \omega + \tilde{a} \right) \Gamma\left(\frac{1}{2} - i\frac{\omega - m\Omega}{2\pi T_{H}} + 2 i M \omega + \tilde{a} \right) \Gamma\left(\frac{1}{2} - 2 i M \omega + \tilde{a} \right)} + \left(\tilde{a} \to -\tilde{a}\right) \right|^{-2}
    \end{split}
\end{equation}
\section{Quantum correction to Hawking radiation from Schwarzian theory}\label{Sec:Quantum}

\begin{figure}
\begin{tikzpicture}
		\draw[red,thick] (0,6) .. controls (5,6) and (10,6.5) .. (15,8);
		\draw[red,thick] (0,4) .. controls (5,4) and (10,3.5) .. (15,2);
		\draw[green,thick] (4,6.1)--(4.3,5.6)--(3.8,5.2)--(4.2,4.5)--(4,3.9);
		\draw[green,thick] (8,6.5)--(8.3,6.2)--(7.8,5.6)--(8.2,5.1)--(7.7,4.5)--(8.2,3.9)--(8,3.5);
 		\node at (8.1,5){\huge{$\phi_{0}$}};
		\draw [decorate,decoration={brace,amplitude=10pt}] (0,7)--(8,7) node[yshift=20pt,midway,brown]{\huge{Near-horizon Region}};
		\draw [decorate,decoration={brace,amplitude=10pt,mirror}] (4,1.5)--(15,1.5) node[yshift=-20pt,midway,brown]{\huge{Far Horizon Region}};
		\draw [decorate,decoration={brace,amplitude=10pt,mirror}] (4,3)--(8,3) node[yshift=-20pt,midway,brown]{\huge{Overlapping Region}};
		\draw [<-,thick] (8,6.8)--(9,8.5) node[yshift=4pt,xshift=4pt,blue]{\large{Quantum Fluctuations at the Boundary}};
		\draw[red,thick] (0,4)--(0,6);
		\draw [<-,thick] (0.2,4.8)--(1,2.8) node[xshift=4pt,yshift=-5pt,blue]{\large{Killing horizon}};
		\draw [<-,thick] (14,6.2)--(13,9) node[yshift=5pt,blue]{\large{Asymptotically Flat Region}};
		\draw[<-,thick] (13,5.3)--(14,5.3) node[yshift=5pt,xshift=-5pt]{\large{Ingoing}};
		\draw[<-,thick] (14,4.7)--(13,4.7) node[yshift=-5pt,xshift=20pt]{\large{Outgoing}};
		\draw[<-,thick] (0.2,5)--(1.2,5) node[yshift=5pt,xshift=-9pt]{\large{Absorbed}};
		\draw[decorate, decoration={snake}, thick,green] (0.5, 4.3) -- (1.5, 4.3);
		\draw[decorate, decoration={snake}, thick,green] (2.5, 5.5) -- (3.5, 5.5);
		\draw[decorate, decoration={snake}, thick,green] (4.3, 5.2) -- (5.6, 5.2);
		\draw[decorate, decoration={snake}, thick,green] (6.2, 5.8) -- (7.3, 5.8);
		\draw[decorate, decoration={snake}, thick,green] (6.6, 4.0) -- (7.7, 4.0);
	\end{tikzpicture}
 \caption{Schematic depiction of the black hole absorption process leading to Hawking radiation as spontaneous emission. We emphasize the role of quantum fluctuations in green.}
\end{figure}

 \subsection{The low-energy effective theory in the throat}

Extremal black holes exhibit an infinitely long AdS$_{2}$ throat near their horizon \cite{Kunduri:2013gce}. Standard dimensional reduction arguments suggest that the two-dimensional theory on AdS$_{2}$ is Jackiw-Teitelboim (JT) gravity with a negative cosmological constant \cite{Jackiw:1984je, Teitelboim:1983ux, Almheiri:2014cka}
\begin{equation} \label{eq:JT_action_0}
    I_{\text{JT}} = -\frac{1}{2}\int_{M} d^2x\ \sqrt{g^{(2)}}\ \phi \left(R + \frac{2}{L^2_{\text{AdS}_2}} \right) - \phi_{b} \int_{\partial M} du\sqrt{h} \left(K - \frac{1}{L^2_{\text{AdS}_2}} \right).
\end{equation}
We note that the derivation of an effective two-dimensional theory, specifically Jackiw-Teitelboim (JT) gravity, from the near-horizon geometry of rotating black holes has been a subject of discussion in the literature \cite{Moitra:2019bub, Castro:2019crn}. The dimensional reduction of rotating black holes to a two-dimensional effective theory is more subtle than its spherically symmetric counterpart due to the breaking of spherical symmetry. It has been pointed out that certain ansatze, such as the one in \cite{Moitra:2019bub}. for the metric perturbations in the decoupling limit may not fully satisfy the linearized Einstein equations or reconstruct the full Kerr geometry \cite{Castro:2019crn}. A more complete treatment, as detailed in \cite{Castro:2019crn}, requires a consistent ansatz that incorporates additional fields and warp factors\footnote{We thank the referee for this insight.}. While a complete consensus on the precise form of the 2D effective action is still emerging, the JT description has proven remarkably successful in capturing key thermodynamic and chaotic properties of near-extremal black holes. In this work, we proceed within this established JT framework.

Since there is no bulk dynamics in two-dimensional gravity, the entire dynamics in \eqref{eq:JT_action_0} arises from the boundary term, which is equivalent to a Schwarzian action
\begin{equation}
    I_{\text{Sch}} \sim \int_{0}^{\frac{1}{T_{H}}} du \{\tan\left(\pi\, T_{H} f(u) \right), u \},\quad \{g(u), u\} \equiv \frac{g'''}{g'} - \frac{3}{2} \left(\frac{g''}{g'} \right)^2,
\end{equation}
the same action happens to govern the low energy dynamics of SYK model \cite{Sachdev:1992fk, Kitaev}, thus leading to an instructive example of lower dimensional holographic duality.

The higher dimensional black hole usually carries conserved charges such as an electric charge or a conserved angular momentum. If we are interested in a grand canonical description of the system near extremality, where fluctuations in these conserved quantitites are allowed, there usually appears an additional $U(1)$ or phase mode in the lower dimensional theory \cite{Castro:2018ffi,Moitra:2018jqs, Sachdev:2019bjn,Moitra:2019bub}. The effective action of lower dimensional JT gravity coupled to a $U(1)$ mode is equivalent to the low energy action of a complex SYK model with global $U(1)$ symmetry \cite{Davison:2016ngz, Sachdev:2019bjn}
\begin{equation} \label{eq:complex_SYK_action}
    I_{Sch\times U(1)} = -C \int_{0}^{\frac{1}{T_{H}}} d\tau \{\tan\left(\pi\, T_{H} f\left(\tau\right)\right), \tau\} + \frac{K}{2} \int_{0}^{\frac{1}{T_{H}}} d\tau \left(\partial{\tau}\phi - i \left(2\pi\mathcal{E}T_{H}\right) \partial_{\tau} f\right)^2.
\end{equation}
The action contains three parameters: $C$, $K$, and $\mathcal{E}$; they are specified by their connection to the thermodynamics of the original four dimensional black hole \cite{Sachdev:2019bjn}
\begin{equation}
    C = \frac{1}{4\pi^2} \left(\frac{\partial S}{\partial T}\right)_{T_{H} \to 0}, \quad \mathcal{E} = \frac{1}{2\pi} \left(\frac{\partial S}{\partial Q}\right)_{T_{H} \to 0}, \quad K = \left(\frac{d Q}{d \mu}\right)_{T_{H} \to 0}.
\end{equation}
$K$ is often dubbed the charge susceptibility or compressibility of the four dimensional black hole. For a Kerr black hole, the role of the charge $Q$, and chemical potential $\mu$ are played by the angular momentum $J = M a$, and the angular velocity $\Omega$. The parameters of a Kerr black hole can be expressed in terms of its horizon radii $r_{+}$ and $r_{-}$ as
\begin{equation}
    M = \frac{r_{+}+r_{-}}{2},\quad a = \sqrt{r_{+} r_{-}},\quad T = \frac{r_{+} - r_{-}}{4\pi r_{+} \left(r_{+} + r_{-}\right)},\quad \Omega = \frac{\sqrt{r_{+} r_{-}}}{r_{+}\left(r_{+} + r_{-}\right)}.
\end{equation}
In an ensemble where the angular momentum $J$ is fixed at $J = r_{0}^2$, the thermodynamic entropy admits a low temperature expansion \eqref{eq:entropy_lowT_canonical}
\begin{equation}
    S = 2 \pi M r_{+} \simeq 2\pi r_{0}^2 + 8\pi^2 r_{0}^3\, T_{H},
\end{equation}
where $r_{0}$ is the event horizon at extremality. From this relationship we obtain
\begin{equation}
    C = 2 r_{0}^3,\quad \mathcal{E} = 1.
\end{equation}
The charge susceptibility is similarly given by
\begin{equation}
    K = - 4 r_{0}^3.
\end{equation}
Therefore, we get that the effective 1D action responsible for the description of the near-horizon dynamics of a Kerr black hole near extremality is
\begin{equation} \label{eq:JT_action_Kerr}
    I_{\text{Sch} \times U(1)} = -2 r_{0}^3 \int_{0}^{\frac{1}{T_{H}}} d\tau \{\tan\left(\pi\, T_{H} f\left(\tau\right)\right), \tau\} - 2 r_{0}^3 \int_{0}^{\frac{1}{T_{H}}} d\tau \left(\partial{\tau}\phi - i \left(2\pi T_{H}\right) \partial_{\tau} f\right)^2. 
\end{equation}

When the black hole has an angular momentum $(J)$ as well as an electric charge $(Q)$, such as those in the Kerr-Newman family, the effective one-dimensional action receives contributions from two $U(1)$ modes
\begin{equation} \label{eq:JT_action_KN}
\begin{split}
    I_{\text{eff}} = -C \int_{0}^{\frac{1}{T_{H}}} d\tau \{\tan\left(\pi T_{H} f\left(\tau\right)\right), \tau\} &+ \frac{K_{e}}{2} \int_{0}^{\frac{1}{T_{H}}} d\tau \left(\partial{\tau}\phi_{e} - i \left(2\pi \mathcal{E}_{e} T_{H} \right) \partial_{\tau} f\right)^2\\ &+ \frac{K_{r}}{2} \int_{0}^{\frac{1}{T_{H}}} d\tau \left(\partial{\tau}\phi_{r} - i \left(2\pi \mathcal{E}_{r} T_{H} \right) \partial_{\tau} f\right)^2,
\end{split}
\end{equation}
and thermodynamics once again determines all the parameters
\begin{subequations} \label{eq:thermodynamic_parameters_KN}
    \begin{align}
    C &= \frac{1}{4\pi^{2}} \left.\left(\frac{\partial S}{\partial T}\right)_{Q,J}\right|_{T_{H} \to 0} = \sqrt{a^{2}+Q^{2}}(2a^{2}+Q^{2}),\\
    K_{e} &= \left.\left(\frac{\partial Q}{\partial \mu_{e}}\right)_{\Omega_{H}}\right|_{T_{H} \to 0} = \frac{\sqrt{a^{2}+Q^{2}}(Q^{2}+2a^{2})(2a^{2}-Q^{2})}{a^{2}(2a^{2}+3Q^{2})},\\
    K_{r} &= \left.\left(\frac{\partial J}{\partial \Omega_{H}}\right)_{\mu_{e}}\right|_{T_{H}\to 0} = - 2\sqrt{a^{2}+Q^{2}}(2a^{2}+Q^{2}),\\    	
    \mathcal{E}_{e} &= \frac{1}{2\pi} \left.\left(\frac{\partial S}{\partial Q}\right)_{J}\right|_{T_{H}\to 0} = \frac{Q^{3}}{2a^{2}+Q^{2}},\\
    \mathcal{E}_{r} &= \frac{1}{2 pi} \left.\left(\frac{\partial S}{\partial J}\right)_{Q}\right|_{T_{H} \to 0} = \frac{2 a\sqrt{a^{2}+Q^{2}}}{{2a^{2}+Q^{2}}}.
    \end{align}
\end{subequations}

The expressions on the r.h.s. are understood to have been evaluated on the extremal configuration. Let us note that for a Reissner-Nordstr\"{o}m black hole, $K_{e} \to \infty$; this means the phase mode gets decoupled from the Schwarzian action, and does not influence low temperature quantum corrections to the gravitational path integral \cite{Kolanowski:2024zrq}.

\subsection{Density of states and correlation function}

To calculate the quantum-corrected Hawking emission rate from rotating black holes, we need to know the density of black hole states with a fixed temperature $T_{H}$ and angular velocity $\Omega_{H}$ slightly above extremality, as well as the two-point function of the boundary operator $\mathcal{O}$ in JT gravity coupled to a U(1) mode \eqref{eq:JT_action_Kerr} or \eqref{eq:JT_action_KN}. We now briefly discuss how to obtain these based on the previous works \cite{Mertens:2019tcm, Iliesiu:2020qvm, Iliesiu:2022onk}, which the reader may consult for more rigorous details.

The partition function obtained from the 2D action \eqref{eq:complex_SYK_action} is given by \cite{Mertens:2019tcm}
\begin{equation}
	Z \left(\beta_{H},\mu \right) = \underbrace{e^{S_{0}} \left(\frac{2\pi C}{\beta_{H}} \right)^{\frac{3}{2}} e^{\frac{2\pi^{2} C}{\beta_{H}}}}_{\text{Schwarzian sector}} \underbrace{\sum_{n} e^{-\beta_{H} \left(\frac{n^{2}}{2K} - \mu n \right)}}_{\text{$U(1)$ sector}},
\end{equation}
where $S_{0}$ is the zeroth order or extremal Bekenstein-Hawking entropy, $\beta_{H} = \frac{1}{T_{H}}$ is the inverse Hawking temperature, and $\mu$ is the relevant chemical potential which for Kerr black hole matches with the angular velocity $\Omega_{H}$. Meanwhile, according to the definition of partition function
	\begin{equation}
		Z \left(\beta_{H}, \mu \right) = \sum_{q} \int dE\ \rho (E, q)\ e^{-\beta_{H} E - \beta_{H} \,\mu\, q},
	\end{equation}
where $\rho\left(E, q\right)$ represents the density of states with energy $E$ and charge $q$. It is understood that $q$ is the conserved charge conjugate to the chemical potential $\mu$ in the partition function, depending on the context it could be the electric charge or the angular momentum of the black hole. Writing $\mu_{1} \equiv \mu \beta_{H}$, we obtain 
\begin{equation}
	e^{S_{0}} \left(\frac{2\pi C}{\beta_{H}} \right)^{\frac{3}{2}} e^{\frac{2\pi^{2} C}{\beta_{H}}} \sum_{q} e^{-\beta_{H} \frac{q^{2}}{2K} - \mu_{1} q } = \sum_{q} \int dE\ \rho (E, q)\ e^{-\beta_{H} E - \mu_{1} q},
\end{equation}
Therefore, we have an inverse Laplace transform of the Schwarzian or SL$\left(2, \mathbb{R}\right)$ part of the partition function
\begin{equation}
	2 C\ e^{S_{0}} \sum_{q} e^{\mu_{1} q} \sinh \left(2\pi \sqrt{2 C \left(E-\frac{q^{2}}{2K}\right)} \right) = \sum_{q}\rho(E,q)\ e^{-\mu_{1}q}.
\end{equation}
Now we consider that both LHS and RHS as a series in $\mu_{1}$, and equate the terms order by order to obtain
\begin{equation}\label{Eq:Density}
	\rho \left(E,q \right) = 2 C\ e^{S_{0}} \sinh \left(2\pi \sqrt{2 C \left(E-\frac{q^{2}}{2K} \right)}\right).
\end{equation}
If $K = \infty$, which means the phase mode can be ignored, the density of states becomes the usual density of states in JT gravity.\\
	
To compute the emission rate, we also require the matrix element of an operator $\mathcal{O}$ on the AdS$_2$ boundary between the initial and final energy eigenstates - which are the states of the black hole and radiation before and after emission. This can be calculated by the Schwarzian description of JT gravity. For a boundary operator of conformal dimension $\Delta$ and charge $Q$ this is given by \cite{Mertens:2019tcm}
\begin{equation}
    \begin{split}
    	\rho\left(E, q\right) \left| \bra{f} \mathcal{O} \ket{i} \right|^2 = \frac{C}{\pi^{2}}\ \delta_{q_{1}, q_{2}+q}\ &\frac{\sinh \left(2\pi\sqrt{2C \left(E_{i}-\frac{q_{1}^{2}}{2K} \right)} \right) \sinh \left(2\pi\sqrt{2C \left(E_{f} - \frac{q_{2}^{2}}{2K} \right)} \right)}{\sinh \left(2\pi\sqrt{2C E_{1}} \right)}
    	\\& \times \frac{ \Gamma\left(\Delta \pm i \sqrt{2C\left(E_{i} - \frac{q_{1}^{2}}{2K} \right) } \pm i \sqrt{2C\left(E_{f} - \frac{q_{2}^{2}}{2K} \right) } \right)}{\left(2C \right)^{2\Delta}\ \Gamma\left(2\Delta \right)}.
    \end{split}
\end{equation}
When the black hole carries more than one charge, say $q$ and $p$ (e.g. Kerr-Newman), the correlation function is obtained by replacing $E - \frac{q^2}{2 K} \rightarrow E - \frac{q^2}{2 K_q} - \frac{p^2}{2 K_p}$.    
\subsection{Hawking radiation corrected by JT gravity fluctuations}

We calculate Hawking emission rate from a Schwarzian two-point function using Fermi's golden rule following the logic of \cite{Brown:2024ajk} (see also \cite{Das:1996we, Maldacena:1997ih, Gubser:1998bc} for an effective string approach to calculate semiclassical greybody factor, which inspired the latest analyses). In pure JT gravity coupled to matter, there is no spontaneous emission because all quanta are reflected back and never reach infinity. Stimulated emission with frequency $\omega$ may occur if a classical external source which oscillates with the same frequency is turned on. The perturbation fields discussed in the previous section play the role of such classical source.

Consider a scalar field for example. Asymptotically, solutions to the scalar wave equation in AdS$_2$ take the form
\begin{equation}
    \phi(t, z) = \phi_{\text{bdy}}\ z^{1 - \Delta} + O\left(z^{\Delta}\right),
\end{equation}
where $\Delta$ is the scaling dimension of the dual primary operator $\mathcal{O}$ in the one-dimensional theory. The coefficient $\phi_{\text{bdy}}$ acts as a source of this operator and couples it to the effective theory on the throat
\begin{equation}
    I = I_{\text{Sch} \times U(1)} + \int dt\ \phi_{\text{bdy}}(t) \mathcal{O}(t).
\end{equation}
In the absence of an external incoming wave, the classical value $\phi_{\text{bdy}}(t)$ is zero. Nonetheless, upon quantization the corresponding operator $\phi_{\text{bdy}}(t) \to \hat{\phi}_{0}(t)$ acts non-trivially on the far-field Hilbert space. Then, the interaction Hamiltonian \[H_{I}(t) = \mathcal{O}(t) \hat{\phi}_{0}(t),\] can be thought of as a coupling between near- and far-region Hilbert spaces. Since the far-region fields are assumed to be free, one may still use the classical e.o.m. to relate $\hat{\phi}_{0}$ to a linear combination of the far-region creation and annihilation operators \[\hat{\phi}_{0} \propto \int_{0}^{\infty} d\omega\, \phi_{0}(\omega) \left(a(\omega) + a^{\dagger}(\omega) \right).\]
Therefore, one expects to obtain a spontaneous transition rate
\begin{equation}\label{Eq:Gamma_i_f}
\begin{split}
    \Gamma_{i \to f} &=  \left|\phi_{0}\right|^2 \left|\langle E_{f}, \omega\right| \left.\mathcal{O}(0) \right| \left. E_{i} \rangle \right|^2\, \delta \left(E_{i} - E_{f} - \omega \right).
\end{split}
\end{equation}
Here $\ket{E_{i}} = \ket{E_{i}} \otimes \ket{\Omega}$ is a state where the black hole has energy $E_{i}$ and there is no Hawking quanta in the far-region, hence the far-field modes are in their vacuum state $\ket{\Omega}$ such that $a(\omega)\ket{\Omega} = 0$. The black hole then emits a quanta of frequency $\omega$ to go to a state with energy $E_{f}$ -- this is the final state $\ket{E_{f}, \omega} = \ket{E_{f}} \otimes a^{\dagger}(\omega) \ket{\Omega}$.

The full spontaneous emission rates is obtained by multiplying with the density of states, and integrating over all possible frequencies $\omega$ and energy $E_{f}$
\begin{equation}
    \Gamma_{\text{spon.}} = \int d\omega\,\omega \int dE_{f}\, \rho\left(E_{f}, \omega\right)\,\Gamma_{i \to f}.
\end{equation}
The density of states of the far-field quanta are always taken to be unity, and the density of states of Schwarzian theory is given in \eqref{Eq:Density}. The rate of change of energy is calculated by
\begin{equation}\label{Eq:rate}
    \frac{dE}{dt\, d\omega } =  \omega\int dE_{f}\, \rho\left(E_{f}\right)\, \Gamma_{i \to f}. 
\end{equation}

\section{Massless scalar emission from a Kerr-Newman black hole}\label{Sec:Scalar}

Recall that the range of applicability of the Teukolksy equation is partially limited. The equation can appropriately describe a scalar field in the general Kerr-Newman black hole. However, for photons and gravitons it only applies in the Kerr black hole. Accordingly, we will discuss the quantum-corrected Hawking emission in two separate sections, in this section we discuss a scalar field in the general Kerr-Newman black hole and in Section \ref{Sec:PhotonGraviton} we address quantum-corrected emission of photons and gravitons in Kerr black holes.

We start our calculation of quantum-corrected Hawking radiation from a Kerr-Newman black hole with the simplest possible example of a massless, spin zero scalar field of charge $e$. We have already described in fair detail how to decompose a field propagating in a Kerr background while maintaining the axial symmetry of the spacetime geometry, and obtain the Teukolsky perturbation equations \eqref{eq:Teukolsky_radial-1} and \eqref{eq:Teukolsky_angular-1} in section \ref{sec:Teukolsky_master_equations}. When the background spacetime is Kerr-Newman, a similar approach can be taken at least for a spin zero field. For the convenience of the reader, let us briefly explain how this is done. We use a radial mode $R_{0}(r)$ and spin-weighted spheroidal harmonics $S_{\ell m}\left(\theta\right)$ to decompose the field as
\begin{equation}
    \Phi\left(t, r, \theta, \phi\right) = e^{-i\omega t + im\phi}\ S_{\ell m}\left(\theta\right) R_{0} \left(r\right),
\end{equation}
 which obeys the following radial and angular Teukolsky equations
 \begin{subequations}
 \begin{align}
    \Delta \frac{d^2R_{0}}{dr^2} + \frac{d\Delta}{dr} \frac{dR_{0}}{dr} + \left(\frac{\left(\left(r^2+a^2\right)\omega - a m - e Q r \right)^2}{\Delta} + 2am\omega - K_{\ell} \right) R_{0} &= 0, \label{eq:Teukolsky_radial} \\
    \partial_{z} \left(1-z^2\right)\partial_{z}S_{\ell m} + \left(K_{\ell} - a^2\omega^2\left(1-z^2\right) - \frac{m^2}{1-z^2} \right) S_{\ell m} &= 0, \label{eq:Teukolsky_angular}
\end{align}
\end{subequations}
where $z = \cos\theta$. To solve the radial equation, we introduce a new variable
\begin{equation*}
    x = \frac{r - r_{+}}{r_{+} - r_{-}},\quad x \in \left[0, \infty \right).
\end{equation*}
The radial equation in this new variable is given by
\begin{equation}
    x\left(x+1\right)\frac{d^2 R_{0}}{dx^2} + \left(2 x + 1\right) \frac{dR_{0}}{dx} + \left(\frac{\mathcal{V}(x)}{x\left(x+1\right)\left(r_{+}-r_{-}\right)^2} + 2 a m \omega - K_{\ell} \right) R_{0} = 0,
\end{equation}
with
\begin{equation}
    \mathcal{V}(x) = \left( \left( \left( \left(r_{+} - r_{-} \right) x + r_{+} \right)^2 + a^2 \right)\omega - a m -eQ\left( \left( r_{+}-r_{-}\right) x+r_{+}\right) \right) ^2.
\end{equation}
In the near horizon region $\left(x \ll 1\right)$, the equation is well approximated by
\begin{equation} \label{eq:scalar_near_equation}
    x\left(x+1\right)\frac{d^2 R_{0}}{dx^2} + \left(2 x + 1\right) \frac{dR_{0}}{dx} + \left(\frac{p_{1}x^{2}+p_{2}x+p_{3}}{x\left(x+1\right)} - p_{4} \right) R_{0} = 0,
\end{equation}
with
\begin{equation}
    \begin{split}
    	&p_{1} = e^{2}Q^{2}-6eQr_{+}\omega+2\omega \left( -am+a^{2}\omega+3r_{+}^{2}\omega \right),
    	\\&p_{2} = \frac{1}{\delta} \left(eQ-2r_{+}\omega)(2am+2eQr_{+}-2a^{2}\omega-2r_{+}^{2}\omega \right),
            \\&p_{3} = \frac{\left(2am+2eQr_{+}-2a^{2}\omega-2r_{+}^{2}\omega \right)^{2}}{4\delta^{2}}
            \\&p_{4} = K_{\ell} - 2am\omega,
    \end{split}
\end{equation}
where we introduced $\delta = r_{+} - r_{-}$ for notational simplicity. It may be noted that the near-horizon approximation does not depend on whether $\omega$ is large or small. On the other hand, the far-horizon approximation as was done in \cite{Maldacena:1997ih, Bredberg:2009pv, Hartman:2009nz} requires that $\omega - m\Omega_{H}$ is an $O(1)$ quantity. We do not use the far-horizon approximation in this paper at all; rather we find the field at infinity using insights from the connection formulae of confluent Heun equation, as described in  \cite{Bonelli:2021uvf}. The radial dictionary analogous to \eqref{eq:Heun_radial_dictionary} for the Kerr-Newman black hole is given by 
\begin{equation}
\begin{split}
	E &= \frac{1}{4} + K_{\ell} - e^{2}Q^{2} + 2eQ(2\delta+3r_{-})\omega - 3\delta^{2}\omega^{2}-8\delta r_{-}\omega^{2} - 6 r_{-}^{2}\omega^{2},\\
	a_{1} &= \frac{i}{\delta} \left(am + eQr_{-} - r_{-}^{2} \omega-a^{2}\omega^{2} \right),\\
	a_{2} &= \frac{i}{\delta} \left(am-a^{2}\omega-\delta^{2}\omega + \delta \left(eQ-2r_{-}\omega \right) + r_{-} \left(eQ-r_{-}\omega \right) \right),\\
	m_{3} &= i e Q - i(r_{+} + r_{-})\omega.
\end{split}
\end{equation}

It can be shown that equation \eqref{eq:scalar_near_equation} admits the following solution which obeys ingoing boundary condition at the horizon $\left(x = 0\right)$
\begin{equation} \label{eq:near_scalar_solution}
    \begin{split}
    R_{0}(x) =& N\, A_{1}\, x^{-i \sqrt{p_{3}}} \left(x+1\right)^{i\sqrt{p_{1}-p_{2}+p_{3}}}\times
    \\&{}_{2}F_{1}\left(\frac{1}{2} - i \sqrt{p_{3}}+i\sqrt{p_{1}-p_{2}+p_{3}} - \tilde{\beta}, \frac{1}{2} - i \sqrt{p_{3}}+i\sqrt{p_{1}-p_{2}+p_{3}} + \tilde{\beta}, 1 - 2 i \sqrt{p_{3}}, -x \right),
    \end{split}
\end{equation}
such that
\begin{equation}
    \tilde{\beta}^2 = \frac{1}{4} - p_{1} + p_{4}.
\end{equation}
In equation \eqref{eq:near_scalar_solution}, $A_{1}$ is a constant of integration, and $N$ is a normalization constant. The normalization is fixed by equating the total flux over the two-sphere to unity. Away from the horizon, $R_{0}(x)$ admits the asymptotic expansion
\begin{equation}
\begin{split}
    R_{0}(x) \xrightarrow{x \to \infty} &N\, A_{1} \left( x^{-\frac{1}{2}-\tilde{\beta}} \frac{\Gamma \left(1 - 2 i \sqrt{p_{3}} \right) \Gamma \left(-2 \tilde{\beta} \right)}{\Gamma\left(\frac{1}{2} - i \sqrt{p_{3}} - i \sqrt{p_{1} - p_{2} +p_{3}} - \tilde{\beta} \right) \Gamma\left(\frac{1}{2} - i\sqrt{p_{3}} + i \sqrt{p_{1} - p_{2} +p_{3}} - \tilde{\beta} \right)} \right.\\ & \left. + x^{-\frac{1}{2}+\tilde{\beta}} \frac{\Gamma \left(1 - 2 i \sqrt{p_{3}} \right) \Gamma \left(2 \tilde{\beta} \right)}{\Gamma\left(\frac{1}{2} - i \sqrt{p_{3}} - i \sqrt{p_{1} - p_{2} +p_{3}} + \tilde{\beta} \right) \Gamma\left(\frac{1}{2} - i\sqrt{p_{3}} + i \sqrt{p_{1} - p_{2} +p_{3}} + \tilde{\beta} \right)} \right).
\end{split}
\end{equation}
The asymptotic form of $R_{0}(x)$ matches the expected asymptotic behavior of a massless scalar field which solves the wave equation on AdS$_2$. Therefore, we identify the source $\phi_{0}$ to the boundary operator as the coefficient of the non-normalizable mode which goes as $x^{-\frac{1}{2} + \tilde{\beta}}$
\begin{equation}\label{Eq:phi0}
    \phi_{0} = \alpha N\, A_{1}\, \frac{\Gamma \left(1 - 2 i \sqrt{p_{3}} \right) \Gamma \left(2 \tilde{\beta} \right)}{\Gamma\left(\frac{1}{2} - i \sqrt{p_{3}} - i \sqrt{p_{1} - p_{2} +p_{3}} + \tilde{\beta} \right) \Gamma\left(\frac{1}{2} - i\sqrt{p_{3}} + i \sqrt{p_{1} - p_{2} +p_{3}} + \tilde{\beta} \right)}.
\end{equation}
Here $\alpha$ is an overall normalization constant that only depends on $s$ and $\ell$. For $\left(s, \ell \right) = \left(0,0 \right)$, $\alpha=1$, for $\left(s, \ell \right) = \left(0, 1\right)$, $\alpha^{2}=6$. We multiply this constant in order to match the quantum-corrected emission rate with the emission rate predicted by a semiclassical analysis in the regime $\frac{E_{i}}{E_{\text{brk}}} \gg 1$ and $M\omega \ll 1$ of the parameter space when the influence of Schwarzian modes is expected to be weak. We also identify the conformal dimension of the operator as
\begin{equation}
    \Delta = \frac{1}{2} + \tilde{\beta}.
\end{equation}
Putting together equations \eqref{Eq:Gamma_i_f}, \eqref{Eq:rate} and \eqref{Eq:phi0} we obtain for a neutral scalar particle that,
\begin{equation} \label{eq:quantum_emission_rate__KN_scalar}
    \begin{split}
    	\frac{dE}{dtd\omega} =\alpha^{2} &\frac{\omega}{\pi \left(r_{+}-r_{-} \right)} L_{2}^{4\tilde{\beta}} \times \left|\frac{ \Gamma \left(\frac{1}{2}+m_{3}+\tilde{\beta} \right)}{\Gamma \left(1+2\tilde{\beta} \right) \left(-2i\omega \left(r_{+}-r_{-}\right) \right)^{-\tilde{\beta}+m_{3}}} \right|^{2} \\
    	\\&\times \frac{C}{\pi^{2}} \sinh \left(2\pi\sqrt{2C \left(E_{f} - \frac{j^{2}n_{2}^{2}}{2K_{r}} - \frac{e^{2}}{2K_{e}} \right)} \right) \\ &\times \frac{ \Gamma \left(\frac{1}{2} + \tilde{\beta} \pm i \left(\sqrt{2 C E_{i}} \pm \sqrt{2 C \left(E_{f} - \frac{j^{2}n_{2}^{2}}{2K} - \frac{e^{2}}{2K_{e}} \right)} \right) \right)}{\left(2C \right)^{1+2\tilde{\beta}} \Gamma \left(1+2\tilde{\beta}\right)}.
    \end{split}
\end{equation}
The thermodynamic parameters entering the last equation are tabulated in equation \eqref{eq:thermodynamic_parameters_KN}, with $\mu_{r, e} \equiv 2\pi T_{H} \mathcal{E}_{r, e}$. In evaluating the above expressions, we will limit ourselves to the leading order in $T_H$ contribution to the greybody factor.

\subsection{The Reissner-Nordstrom limit}

If we consider the limit where the Kerr-Newman solution becomes the Reissner-Nordstrom one and further specialize to the case $s=0$ and $\ell=0$, we find
	\begin{eqnarray}
		E&=&\left(\frac{1}{2} + \ell \right)^{2}, \qquad 
	    a_{1}=-\frac{i\omega}{4\pi T_{H}}, \qquad 
    	a_{2}=-\frac{i\omega}{4\pi T_{H}}, \qquad  
    	m_{3}=-2iQ\omega, \nonumber \\
    	p_{1}&=&6\omega^{2}Q^{2}, \qquad 
    	p_{2}=\frac{Q\omega^{2}}{\pi T}, \qquad  
    	p_{3}=\frac{\omega^{2}}{16\pi^{2}T_{H}^{2}}, \qquad     	p_{4}=\ell\left(\ell+1 \right).
    \end{eqnarray}
    
    The final result for the emission is 
    \begin{equation}
    	\frac{dE}{dtd\omega}=\frac{2}{\pi}r_{+}^{2}\omega^{3}\frac{\sinh \left(2C E_{f} \right)}{\sinh \left(2C E_{i} \right)-\sinh \left(2C E_{f} \right)},
    \end{equation}
which precisely reproduces the results recently reported in \cite{Brown:2024ajk} for the quantum-corrected Hawking emission rate of a neutral scalar particle in the Reissner-Nordstr\"om background.

\subsection{The semiclassical limit} \label{SubSec:SemiclassicalLimit-Scalar}

An important sanity check for our computations is to reproduce the semiclassical limit of Hawking radiation, that is, the expression for Hawking emission given by the product of the Bose factor and the greybody factor.  Our goal is to streamline some of the computations by emphasizing certain general properties. We seek to evaluate the difference between the final and initial energy of the black hole, $E_{f}-E_{i}$. Naively, the difference between the initial and finite state energy of the black hole $E_f-E_i$ depends on the frequency of the emitted quantum. In the presence of an electric field, such relation is affected by the way we approach the near-horizon region. We define $\Delta_{t}\chi=\phi_{e}$, where $\phi_{e}$ is the electric potential and therefore, work in the gauge  $\chi=\phi t$ which implies 
\begin{equation}
    	\Phi \rightarrow \Phi e^{ie\chi},
    \end{equation}
	here $\Phi$ is the wave function. The full wave function is then 
	\begin{equation}
		\Phi=e^{-i(\omega-e \phi_{e})t+im\phi}S(\theta)R(r).
	\end{equation}
When reducing the four-dimensional gravity to an effective two-dimensional system, we use a transformed coordinate $\phi^{\prime}=\phi-\Omega_{H}t$ \cite{Moitra:2019bub}, leading to 
    \begin{equation}
        \Phi=e^{-i(\omega-e \phi_{e}-m\Omega_{H})t+im\phi^{\prime}}S(\theta)R(r).
    \end{equation}
Thus, the effective energy under this coordinate transformation is no longer $\omega$, rather, we should use the effective 
	\begin{equation} \label{eq:omega_eff}
		\omega_{\mathrm{eff}}=\omega-m\Omega_{H}-e\phi_{e}.
	\end{equation} 
	
We conclude this short discussion by noting that in the appropriate limit, we always recover the standard formula for the Hawking radiation rate which is given by the product of the Bose factor and the greybody factor. Starting from equation \eqref{eq:quantum_emission_rate__KN_scalar}, we consider the semiclassical limit where $E_{i}\gg E_{\rm brk}$ and $E_{\mathrm{Gap}} \ll E_{i}$. For $m=0$, we can obtain the bosonic factor by making the following approximations
\begin{equation}\label{bosonic_factor}
	\begin{split}
		&\sinh \left(2\pi\sqrt{\frac{E_{i}-\omega_{\mathrm{eff}}}{E_{\mathrm{brk}}}} \right) \Gamma \left(1\pm i\sqrt{\frac{2E_{i}}{E_{\mathrm{brk}}}}   \pm i\sqrt{\frac{2E_{i}-\omega_{\mathrm{eff}}}{E_{\mathrm{brk}}}} \right),\\
        =\ &\sinh \left(2\pi\sqrt{\frac{E_{i}-\omega_{\mathrm{eff}}}{E_{\mathrm{brk}}}} \right)\frac{\pi \left( \sqrt{\frac{2E_{i}}{E_{\mathrm{brk}}}}+\sqrt{\frac{2(E_{i}-\omega_{\mathrm{eff}})}{E_{\mathrm{brk}}}} \right)}   {\sinh \left( \pi \left( \sqrt{\frac{2E_{i}}{E_{\mathrm{brk}}}}+\sqrt{\frac{2 \left(E_{i}-\omega_{\mathrm{eff}} \right)}{E_{\mathrm{brk}}}} \right) \right)}   \frac{\pi \left( \sqrt{\frac{2E_{i}}{E_{\mathrm{brk}}}} - \sqrt{\frac{2 \left(E_{i}-\omega_{\mathrm{eff}} \right)}{E_{\mathrm{brk}}}}\right)}   {\sinh\left(   \pi\left( \sqrt{\frac{2E_{i}}{E_{\mathrm{brk}}}}-\sqrt{\frac{2\left(E_{i}-\omega_{\mathrm{eff}}\right)}{E_{\mathrm{brk}}}}\right) \right)},\\
        \simeq\ &\frac{e^{2\pi\sqrt{\frac{E_{i}-\omega_{\mathrm{eff}}}{E_{\mathrm{brk}}}}}}{e^{2\pi\sqrt{\frac{E_{i}}{E_{\mathrm{brk}}}}}-e^{2\pi\sqrt{\frac{E_{i}-\omega_{\mathrm{eff}}}{E_{\mathrm{brk}}}}}}\times \frac{4\pi^{2}\omega_{\mathrm{eff}}}{E_{\mathrm{brk}}},\\
        \simeq\ &\frac{1}{e^{  \frac{\omega_{\mathrm{eff}}}{T_{H}}   }-1}\times \frac{4\pi^{2}\omega_{\mathrm{eff}}}{E_{\mathrm{brk}}}.
	\end{split}
    \end{equation}
This expression agrees with standard semiclassical approach where the bosonic thermal factor is dressed with the greybody factor to determine the rate of Hawking radiation.

\subsection{Quantum-corrected emission}
In this section, we present several plots in which the quantum-corrected emission is evaluated and compared with the semiclassical results. We consider the following range of frequencies in quantum-corrected theory: $0 \leq \omega \leq \omega_{\rm up}$. Meanwhile, in semiclassical theory, we consider the range of frequencies $0 \leq \omega \leq \infty$. The value $\omega_{\rm up}$ satisfies $\omega_{\rm eff} \left(\omega_{\rm up}\right) = E_{i},$ with $\omega_{\rm eff}$ given by equation \eqref{eq:omega_eff}, so that at this point the final energy of the black hole vanishes, $E_{f}=E_{i}-\omega_{\mathrm{eff}} = 0$. There is another interesting value, $\omega_{\rm low}$ satisfying $\omega_{\rm eff} \left(\omega_{\rm low} \right) = 0$. Recall that only for $\omega>\omega_{\rm low}$ is the bosonic factor positive. This value $\omega_{\rm low}$ precisely agrees with the upper bound of the superradiance window.



\subsubsection{Enhancement of quantum-corrected emission for particles with angular momentum}
    
\begin{figure}[t]
	\centering
		\includegraphics[width=1\textwidth]{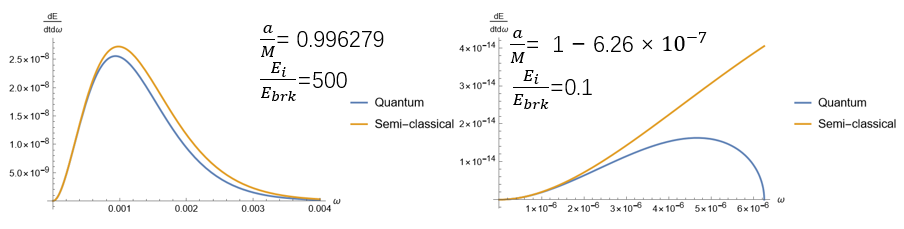}
		\caption{Comparing the emission, $\frac{dE}{dtd\omega}$, in the quantum-corrected and semi-classical framework for a scalar particle with quantum numbers $s=0, \ell=0, m=0$.}
        \label{Fig:Emission-s-m-0}
	\end{figure}

Let us first consider the case of a scalar particle emitted with vanishing angular momentum: $s=0$, $\ell=0$, $m=0$. For $E_{i}=500\,E_{\mathrm{brk}}$, we find that in the quantum-corrected theory, $\frac{dE}{dt}=3.84776\times 10^{-11}$, while in semiclassical theory, $\frac{dE}{dt}=4.34964\times 10^{-11}$. In the strongly coupled regime, $E_{i}=0.1E_{\mathrm{brk}}$, the quantum-corrected theory yields $\frac{dE}{dt}=5.83802\times10^{-20}$, while the  semiclassical theory leads to $\frac{dE}{dt}=1.29904\times10^{-18}$.  The results, shown in Fig. \ref{Fig:Emission-s-m-0}, describe a behavior that is quite similar to the one discussed in the Reissner-Nordst\"om case \cite{Brown:2024ajk}. Namely, we note a generic suppression of the emission due to quantum effects and a qualitative difference that is most pronounced in the quantum regime when the mass of the black hole is smaller than the scale where quantum effects are prominent.

\begin{figure}[t]
		\centering
		\includegraphics[width=1\textwidth]{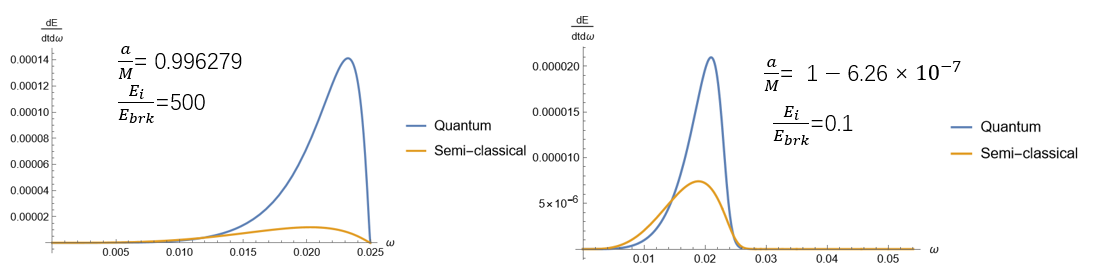}
		\caption{Comparing the emission rate, $\frac{dE}{dtd\omega}$, in quantum-corrected theory and semi-classical theory for a scalar particle emission with positive angular momentum $s=0$, $\ell=1$, $m=1$ for black hole in classical regime (left) and in quantum regime (right).}
        \label{Fig-scalar-m1}
	\end{figure}

Significant qualitative differences arise when we consider particles carrying angular momenta. Let us consider the quantum-corrected emission of scalar particles with angular momentum in the same direction as the black hole, $m=1$. Figure \ref{Fig-scalar-m1} shows a notable enhancement of the quantum-corrected emission with respect to the semiclassical approximation.  For the particular case of $s=0$, $\ell=1$, $m=1$, in the classical regime, we obtain that the quantum-corrected emission is $\frac{dE}{dt}=8.7181\times10^{-7}$, this is represented in the left panel of Fig. \ref{Fig-scalar-m1}; while in the semiclassical theory the result is an order of magnitude smaller, $\frac{dE}{dt}=7.55497\times 10^{-8}$. In the right panel of Fig. \ref{Fig-scalar-m1} we depict the results in the strongly coupled regime. We obtain that the quantum-corrected emission is $\frac{dE}{dt}=7.46214\times10^{-7}$, while in the semiclassical theory   $\frac{dE}{dt}=1.20048\times 10^{-7}$.
Deeper in the quantum regime, where $E_i/E_{\rm brk}\le 1$, we find a more pronounced enhancement of the quantum-corrected emission with respect to the semiclassical result. The emission rates are shown in Fig. \ref{Fig-scalar-m1}. We note that the ratio of quantum-corrected to semiclassical emission rates is about an order of magnitude.

It is worth noting that this enhancement may be linked to a quantum instability. In particular, for some values of the emission frequency $\omega$ that satisfy $\omega_{\rm low} < \omega < \omega_{\rm up}$ (where $\omega_{\rm low}$ is the superradiant threshold), the quantity $\tilde{\beta}$ can become imaginary, signaling a potential instability.

\subsubsection{Emission from Reissner-Nordstr\"om, Kerr-Newman and Kerr black holes}

In the remainder of this section, we compare the emission rates of neutral scalar particles from the three types of black holes: Reissner-Nordst\"om, Kerr-Newman and Kerr black holes. To make this comparison somehow meaningful, we consider parameters of the Kerr-Newman black hole such that all three black holes have the same entropy and the same temperature. More practically, we present the emission rates for the three cases while keeping fixed the combinations  $2a^{2}+Q^{2}$  and $\left(r_{+}-r_{-}\right)$ which function as proxies for the black hole entropy and temperature. In particular, we consider the following numerical values: $S = \pi (2a^{2}+Q^{2})=800\pi$, and $\frac{E_{i}}{E_{\mathrm{brk}}}=500$, fixed for all three black holes.

\begin{figure}[t]
	\centering
    \begin{subfigure}{1\textwidth}
	\includegraphics[width=1\textwidth]{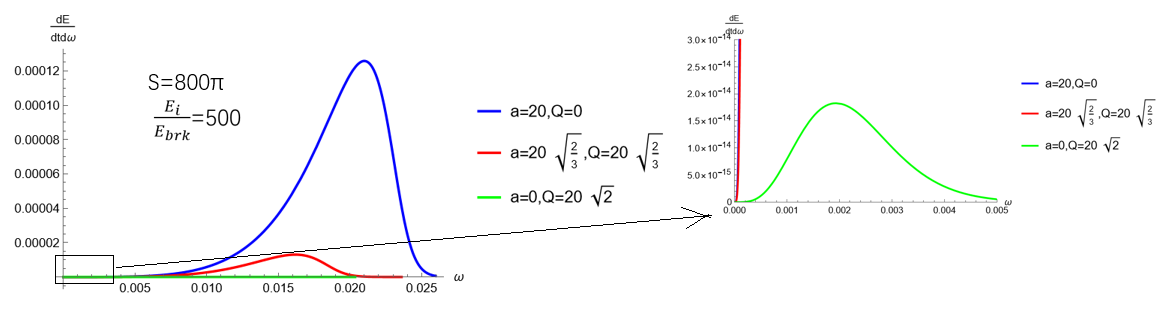}
	\caption{}
    \label{Fig:3BH-highT-Q}
\end{subfigure}
\begin{subfigure}{1\textwidth}
	\centering
	\includegraphics[width=1\textwidth]{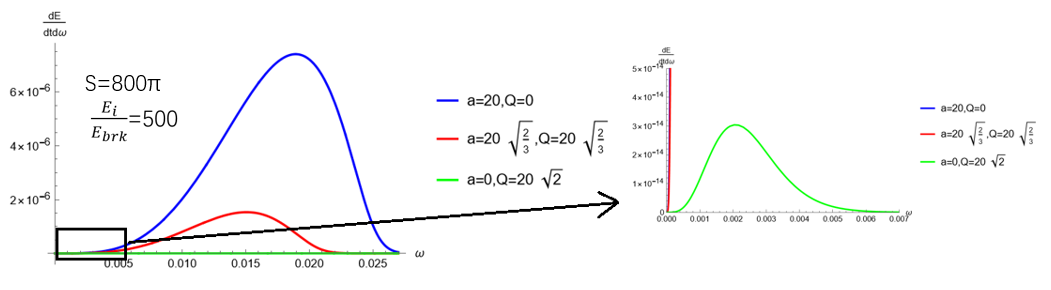}
	\caption{}
    \label{Fig:3BH-highT-SC}
\end{subfigure}
\caption{(a) Quantum-corrected and (b) semiclassical emission rates, $\frac{dE}{dtd\omega}$, for  $s=0$, $\ell=1$, $m=1$ for the three kinds of black holes in the classical regime $\left( E_{i} \gg E_{\rm brk} \right)$. Inset shows the emission rates for RN black hole, which is much suppressed compared to its rotating cousins.}
\end{figure}

Figures \ref{Fig:3BH-highT-Q} and \ref{Fig:3BH-highT-SC} display the emission rates for the quantum-corrected and semiclassical cases, respectively. In both cases the rate of emission is the highest for Kerr black hole. Given that the scalar particles carry angular momenta in the direction of the black hole rotation we see an enhancement in the quantum-corrected case that is two orders of magnitude larger than in the semiclassical case. 

\begin{figure}[t]
  \subcaptionbox*{}[.48\linewidth]{%
    \includegraphics[width=\linewidth]{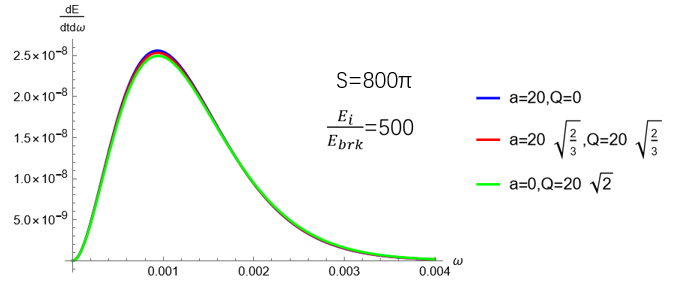}%
  }%
  \hfill
  \subcaptionbox*{}[.48\linewidth]{%
    \includegraphics[width=\linewidth]{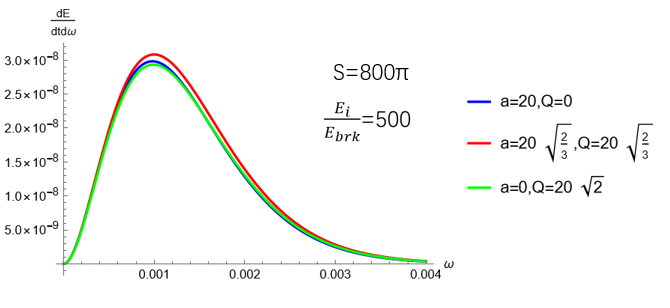}%
  }
  \caption{Emission rates $\frac{dE}{dt\, d\omega}$ are shown for scalar particles with $s=0$, $\ell=0$, and $m=0$. The left panel displays the quantum-corrected emission, while the right panel shows the semiclassical result.}\label{Fig:3BH-highT-m0}
\end{figure}

To highlight that the main difference comes from the fact that the emitted particle has non-vanishing angular momentum, we also plot the emission rates from the three kinds of black holes for scalar particles with vanishing angular momentum. Looking at Fig. \ref{Fig:3BH-highT-m0}, we notice that the rates are essentially indistinguishable. Moreover, the quantum-corrected (left panel) and the semiclassical (right panel) results are of the same order.

\begin{figure}[t]
	\centering
    \begin{subfigure}{1\textwidth}
		\includegraphics[width=1\textwidth]{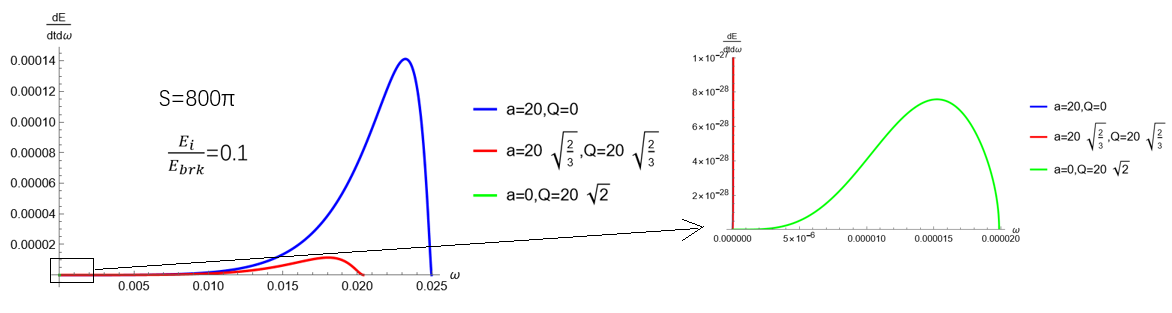}
		\caption{}
        \label{Fig:3BH-lowT-Q}
	\end{subfigure}
    \begin{subfigure}{1\textwidth}
		\centering
		\includegraphics[width=1\textwidth]{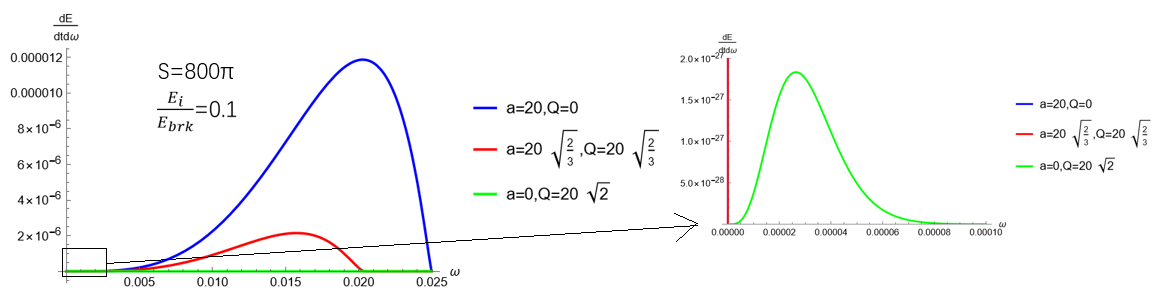}
		\caption{}
        \label{Fig:3BH-lowT-SC}
	\end{subfigure}
    \caption{Emission rates, $\frac{dE}{dtd\omega}$, for  $s=0$, $\ell=1$, $m=1$ for all three kinds of black holes in the strongly coupled regime in (a) quantum-corrected, and (b) semiclassical theory. The emission rate of Reissner-Nordstr\"{o}m black hole is significantly suppressed compared to the other two and is shown in the inset.}
\end{figure}

Next, we consider the following numerical values: $S=\pi (2a^{2}+Q^{2})=800\pi$, and $\frac{E_{i}}{E_{\mathrm{brk}}}=0.1$, fixed for all three black holes. Figures \ref{Fig:3BH-lowT-Q} and \ref{Fig:3BH-lowT-SC} represent the rates calculated using quantum-corrected and semiclassical methods, respectively, in the strongly quantum regime. The results still point to Kerr black hole as the one with maximum rate of radiation.
\begin{figure}[t]
    \centering
    \includegraphics[width=1\textwidth]{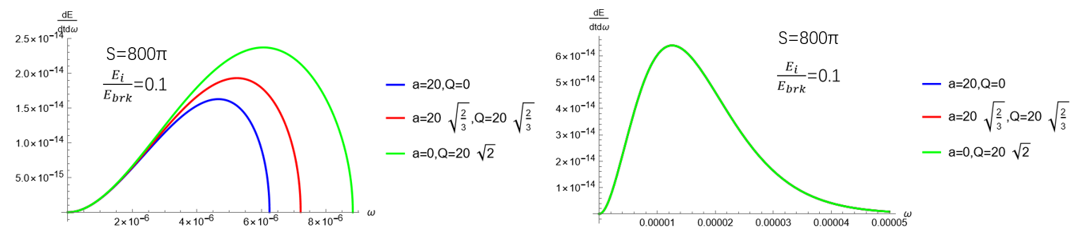}
    \caption{Emission rates, $\frac{dE}{dt\, d\omega}$ for scalar particles carrying no angular momentum, $s=0$, $\ell=0$, $m=0$. The left panel represents the quantum-corrected emission and the right panel the semiclassical one in the strongly coupled regime.}
    \label{Fig:3BH-lowT-m0}
\end{figure}
In the absence of particles carrying angular momenta, Fig. \ref{Fig:3BH-lowT-m0} shows that there is no significant difference among the emission rates of different black holes. In this strongly coupled regime, the Reissner-Nordst\"om black hole has the highest emission rate for particles that do not carry any angular momentum. 

\section{Comments on the emission of photons and gravitons}\label{Sec:PhotonGraviton}

We follow the same strategy as in section \ref{Sec:Scalar} to identify the source for fields with spin $s>0$. In this case, one needs to solve the set of Teukolsky equations \eqref{eq:Teukolsky_radial-1} and \eqref{eq:Teukolsky_angular-1}. Once again, introducing a new variable $x = \frac{r - r_{+}}{r_{+} - r_{-}}$, we rewrite the radial equation as
\begin{equation}
\begin{split}
    x (x+1) &\frac{d^2 R_{s}}{dx^2} + (s+1) \left(2 x + 1\right) \frac{d R_{s}}{dx} \\ &+ \left(\frac{\mathcal{V}(x)}{x (x+1) \left(r_{+} - r_{-} \right)^2} + 2 a m \omega - K_{\ell} + 4 i s \omega \left(r_{+} + \left(r_{+} - r_{-}\right)x\right) \right) R_{s} = 0,
\end{split}
\end{equation}
with
\begin{equation}
\begin{split}
    \mathcal{V}(x) &= K(x)^2 - 2 i s K(x) \left(\left(r_{+}-r_{-}\right)x + r_{+} - M\right),\\
    \text{where}\quad K(x) &= \left( \left( \left(r_{+} - r_{-} \right) x + r_{+} \right)^2 + a^2 \right)\omega - a m.
\end{split}
\end{equation}
As before, we may ignore the terms of $O\left(x^2\right)$ in the near-horizon region. In addition, defining $ \rho_{s} = \Delta^{\frac{s}{2}} R_{s}$ we obtain a simpler equation of motion
\begin{equation}
    	x(x+1)\frac{d^2\rho_{s}}{dx^2} + (2x+1) \frac{d\rho_{s}}{dx} + \left(\frac{p_{1} x^2 + p_{2} x + p_{3}}{x(x+1)} - p_{4} \right) \rho_{s} = 0,
    \end{equation}
where the parameters are defined as
    \begin{align}
    \begin{split}
    	p_{1} &= - s^{2} - 4 i s r_{+}\omega + 2\omega \left( \left(3 r_{+}^2 + a^2 \right)\omega - a m \right),\\
        p_{2} &= \frac{\left(is - 2r_{+}\omega \right)\left( 2 a m + i \left(r_{+} - r_{-}\right) s - 2 \left(r_{+}^{2} + a^{3} \right)\omega \right)}{r_{+} - r_{-}},\\
    	p_{3} &= \frac{\left( 2am + i \left(r_{+} - r_{-}\right) s - 2 \left(a^{2} + r_{+}^{2} \right)\omega \right)^{2}}{4 \left(r_{+} - r_{-}\right)^{2}},\\
    	p_{4} &= K_{\ell} - 2 a m \omega + s - 4 i s r_{+} \omega .
    \end{split}
    \end{align}
Other relevant parameters are
\begin{equation}
    \begin{split}
        E=&\frac{1}{4}+s(s+1)+K_{\ell}-6M^{2}\omega^{2}, \\ 
        a_{1}=&-i\frac{\omega-m\Omega_{H}}{4\pi T_{H}}+2iM\omega+\frac{s}{2}, \\
        a_{2}=&-i\frac{\omega-m\Omega_{H}}{4\pi T_{H}}-\frac{s}{2}, \\
        m_{3}=&-2iM\omega+s,
    \end{split}
\end{equation}
which is the same as \cite{Bonelli:2021uvf}, and we only keep the terms occurring at the leading order in temperature. A solution satisfying the ingoing boundary condition near the horizon is given by
    \begin{equation}
    	\begin{split}
    		R_{s} = N\, A_{1}\, &x^{-\frac{s}{2}-i\sqrt{p_{3}}}\ (x+1)^{-\frac{s}{2} + i\sqrt{p_{3} - p_{2} + p_{1}}} \times \\ & _{2}F_{1} \left(\frac{1}{2}-\frac{1}{2}\sqrt{1 - 4 p_{1} + 4 p_{4}}-i\sqrt{p_{3}} + i\sqrt{p_{3} - p_{2} + p_{1}},\right. \\ & \hspace{4em} \left. \frac{1}{2} + \frac{1}{2} \sqrt{1 - 4p_{1} + 4p_{4}}-i\sqrt{p_{3}}+i\sqrt{p_{3} - p_{2} + p_{1}}, 1-2i\sqrt{p_{3}}, -x \right).
    	\end{split}
    \end{equation}

For large $x$, i.e. near the boundary of AdS$_2$, the asymptotic behavior of this solution is given by
\begin{equation}
    Y_{s} \xrightarrow{x \to \infty} N\, A_{1} \left(x^{-\frac{1}{2} - \frac{1}{2}\sqrt{1 + 4 p_{4} - 4 p_{1}} }\ \gamma_{+} + x^{-\frac{1}{2} + \frac{1}{2}\sqrt{1 + 4 p_{4} - 4 p_{1}} }\ \gamma_{-} \right),
\end{equation}
where
\begin{align}
    \gamma_{\pm} = \frac{\Gamma(\pm\sqrt{1-4p_{1}+4p_{4}})\Gamma(1-2i\sqrt{p_{3}})}{\Gamma(\frac{1}{2} \pm \frac{1}{2}\sqrt{1-4p_{1}+4p_{4}}-i\sqrt{p_{3}}+i\sqrt{p_{1}-p_{2}+p_{3}})\Gamma(\frac{1}{2}-i\sqrt{p_{2}} \mp \sqrt{p_{1}-p_{2}+p_{3}}-\frac{1}{2}\sqrt{1-4p_{1}+4p_{4}})}.
\end{align}
Once again, we choose the coefficient of the non-normalizable mode to be the source to the boundary operator; thus obtaining
\begin{equation}
    \phi_{0} = \alpha N\, A_{1}\, \gamma_{-}.
\end{equation}

Here we list the values for some relevant pairs of quantum numbers $(s,l)$ which will be used in this section. When $(s,l)=(1,1)$, $\alpha^{2}=\frac{2^{3}\Gamma(4)}{\Gamma(3)^{2}} = 12$, when $(s,l)=(2,2)$, $\alpha^{2}=\frac{2^{4}\Gamma(6)}{\Gamma(5)^{2}} = \frac{10}{3}$. There is still a potential extra factor depending on the spin, $s$, of the particle that could be added to $\phi_0$. This factor is dictated by the asymptotic form of the field $Y=A\, x^{\Delta-1}$, and leads to the conformal dimension: 
\begin{equation}
    \Delta=\frac{1}{2}+\frac{1}{2}\sqrt{1+4p_{4}-4p_{1}}.
\end{equation}

\subsection{Photon emission}
The complete expression for the quantum-corrected emission of particles with $s=1$ is
    
\begin{equation}
    \begin{split}
    		\frac{dE}{dtd\omega} = 24\ &\frac{\omega^{3}(r_{+}-r_{-})^{2-2\tilde{\beta}}}{\pi}L_{2}^{4\tilde{\beta}}\\
            & \times \left| \frac{ \Gamma \left(\frac{1}{2}+m_{3}+\tilde{\beta} \right)}{ \Gamma \left( 1+2\tilde{\beta} \right) \left(-2i\omega \left(r_{+}-r_{-} \right) \right)^{-\tilde{\beta}+m_{3} }} \right|^{2}\\
            & \times \frac{C}{\pi^{2}} \sinh \left(2\pi\sqrt{2C \left(E_{2}-\frac{j^{2}n_{2}^{2}}{2K_{r}}-\frac{e^{2}}{2K_{e}} \right)} \right)\\
            & \times \frac{ \Gamma \left(\frac{1}{2}+\tilde{\beta} \pm i\left(\sqrt{2 C E_{1}} \pm \sqrt{2C \left(E_{2}-\frac{j^{2}n_{2}^{2}}{2K}-\frac{e^{2}}{2K_{e}} \right)} \right) \right)}{\left( 2C \right)^{1+2\tilde{\beta}}\Gamma\left(1+2\tilde{\beta}\right)}.
    \end{split}
\end{equation}

\begin{figure}[t]
		\centering
		\includegraphics[width=1\textwidth]{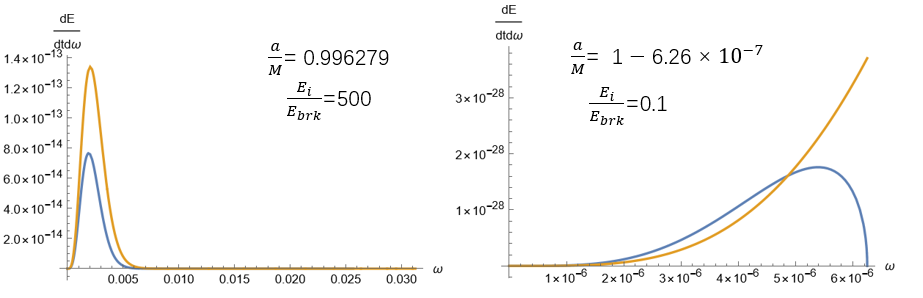}
		\caption{Emission rate $\frac{dE}{dtd\omega}$ in the quantum-corrected theory (blue), and the semiclassical approximation (yellow) for a photon with $s=1$,$\ell=1$,$m=0$. The right panel corresponds to a black hole at a relatively high temperature, and the left panel to a black hole in the quantum regime compared to the quantum scale.}
        \label{Fig:Photon-HighT}
\end{figure}

Let us consider emission of spin 1 particles with vanishing orbital angular momentum, $s=1$, $\ell=1$, $m=0$.  For a black hole in the classical regime $\left( E_{i} \gg E_{\rm brk} \right)$, we find that in the quantum-corrected  theory $\frac{dE}{dt} = 9.53947\times 10^{-17}$, while according to a semiclassical calculation $\frac{dE}{dt} = 3.16125\times 10^{-16}$. Basically, similar to the case of emission from a Reissner-Nordstr\"om black hole \cite{Brown:2024ajk}, we have a suppression of emission as can be seen in the left panel of Fig. \ref{Fig:Photon-HighT}.  For a black hole in the quantum regime $\left(E_{i} \leq E_{\rm brk} \right)$, in quantum-corrected theory $\frac{dE}{dt}=2.29395\times 10^{-34}$, and in semiclassical theory $\frac{dE}{dt}=2.21738\times 10^{-31}$.\\

Teukolsky and Press established in \cite{Teukolsky:1974yv} that the dominant emission channel for photons and gravitons is  $\ell=m=s$.  It is, therefore,  more interesting to explore the quantum number corresponding to the dominant channel for photon emission, $\left(\ell=1, m=1\right)$. Unfortunately, this contribution corresponds to $\tilde{\beta}^2< 0$, which is related to an imaginary conformal dimension for the dual operator. This situation is related to the superradiant bound. Namely, for $\omega=m\,\Omega_H$ we find the issue of $\tilde{\beta}$ becoming imaginary. A similar phenomenon was discussed in \cite{Hartman:2009nz,Bredberg:2009pv}, a complex conformal dimension is clearly an indication of instability. In the spherically symmetric situation, the instability has also been connected to  $Ads_{2}$ Schwinger pair production studied in \cite{Kim:2008xv,Pioline:2005pf}. We will address this instability  in a separate publication.

\subsection{Graviton emission}

The complete expression for the quantum-corrected emission rate of particles with $s=2$ is given by
\begin{equation}
    \begin{split}
    	\frac{dE}{dtd\omega} = \frac{160}{3} &\frac{\omega^{5}(r_{+}-t_{-})^{4-2\tilde{\beta}}}{\pi }L_{2}^{4\tilde{\beta}}\\
        & \times \left| \frac{\Gamma \left(\frac{1}{2}+m_{3}+\tilde{\beta} \right)}{\Gamma \left(1+2\tilde{\beta} \right) \left(-2i\omega \left(r_{+}-r_{-} \right) \right)^{-\sqrt{E}+m_{3} }} \right|^{2}\\
        & \times \frac{C}{\pi^{2}} \sinh \left(2\pi\sqrt{2C \left(E_{2}-\frac{j^{2}n_{2}^{2}}{2K_{r}}-\frac{e^{2}}{2K_{e}} \right)}\right)\\
    	& \times \frac{\Gamma \left(\frac{1}{2}+\tilde{\beta} \pm i \left(\sqrt{2CE_{1}} \pm \sqrt{2C \left(E_{2}-\frac{j^{2}n_{2}^{2}}{2K}-\frac{e^{2}}{2K_{e}} \right)} \right) \right)}{\left(2C \right)^{1+2\tilde{\beta}}\Gamma \left(1+2\tilde{\beta} \right)}.
    \end{split}
\end{equation}

\begin{figure}[t]
	\centering
	\includegraphics[width=1\textwidth]{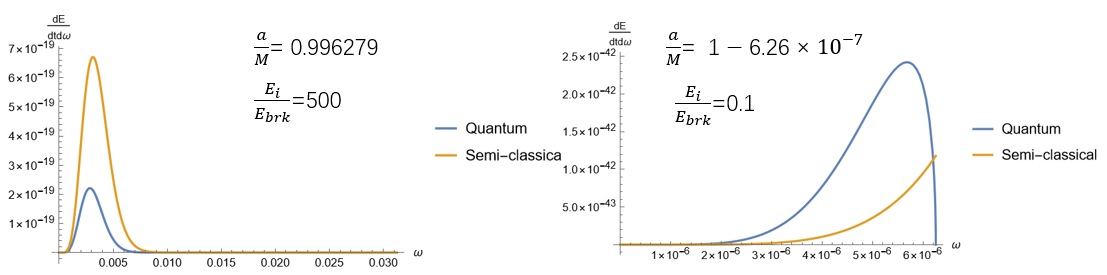}
	\caption{Comparison of emission rates $\frac{dE}{dtd\omega}$ in quantum-corrected and semiclassical theories when $s=2$, $\ell=2$, $m=0$ for a black hole in the classical regime (left) and in the quantum regime (right).}
    \label{Fig:S2m0}
\end{figure}

Let us first address the emission rate when the graviton carries no angular momentum, $s=2$, $\ell=2$, $m=0$.  For black hole in classical regime, in the quantum-corrected theory, $\frac{dE}{dt}=5.70429\times 10^{-22}$ while in the semiclassical approximation the rate is $\frac{dE}{dt}=1.96388\times 10^{-21}$. For black hole in quantum regime, in the quantum-corrected theory, $\frac{dE}{dt}=4.45164\times 10^{-48}$ while in the semiclassical theory, $\frac{dE}{dt}=3.24316\times 10^{-44}$. Fig. \ref{Fig:S2m0}, shows the typical suppression in the quantum-corrected case. 

\begin{figure}[t]
	\centering
	\includegraphics[width=1\textwidth]{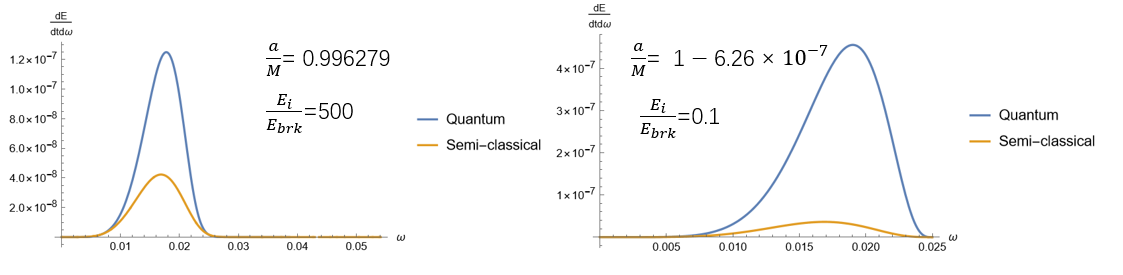}
	\caption{$\frac{dE}{dtd\omega}$ when $s=2$, $\ell=2$,$m=1$ in quantum theory for black hole in classical regime (left) and in quantum regime (right).}
    \label{Fig:s2m1}
\end{figure}

For gravitons that carry angular momentum in the same direction as the black hole, there is a substantial enhancement of the emission. The results for the case of $s=2$, $\ell=2$, $m=1$ are shown in Fig. \ref{Fig:s2m1}. For a black hole in the classical regime, the quantum-corrected emission is $\frac{dE}{dt}=1.01477\times 10^{-9}$, while the semiclassical emission is  $\frac{dE}{dt}=4.0597\times 10^{-10}$. For a black hole in the quantum regime, the emission in the quantum-corrected theory: $\frac{dE}{dt}=3.45649\times 10^{-9}$, while in semiclassical theory: $\frac{dE}{dt}=3.20731\times 10^{-10}$. The quantum enhancement is nearly one order of magnitude.
For these parameters, the semiclassical emission is smaller than the quantum-corrected emission. We could, in fact, estimate the ratio of the emissions. One of the enhancement factors comes from the smaller effective conformal dimension. There are four gamma functions in the correlator, and to estimate the enhancement we only keep the enhancement factor 

\begin{equation}
        k_{{\rm Enhanced}_a}\approx \Gamma(\Delta)^{4}.
\end{equation}
In the expression for $\frac{dE}{dtd\omega}$ there are factors such as $(r_{+}-r_{-})^{\sqrt{1-4p_{1}+4p_{4}}}$,\, $L_{2}^{\sqrt{1-4p_{1}+4p_{4}}}$,\,$(2C)^{\sqrt{1-4p_{1}+4p_{4}}}$, these factors also can be very large so they will contribute to a large factor 
\begin{equation}
    k_{{\rm Enhanced}_b}=(\frac{L_{2}^{2}}{2(r_{+}-r_{-})C})^{\sqrt{1-4p_{1}+4p_{4}}-(2l+1)},
\end{equation}
where we assume $\omega=m\Omega_{H}$
    so the total enhancement could be estimated to approximately  
    \begin{equation}
        k_{{\rm Enhanced}}=k_{{\rm enhance}_a}k_{{\rm Enhance}_b}\approx3.2\times10^{3}
    \end{equation}
We can contrast this estimate with the direct numerical results which yield, for $\omega=m\Omega_{H}$,
    \begin{equation}
        \frac{\frac{dE}{dtd\omega}_{\rm Quantum-Corrected}}{\frac{dE}{dtd\omega}_{{\rm Semi-classical}}}\approx 2.5\times10^{3}.
    \end{equation}

\subsection{The effect of rotation in particle emission and instabilities}

\begin{figure}[t]
	\centering
	\includegraphics[width=0.5\textwidth]{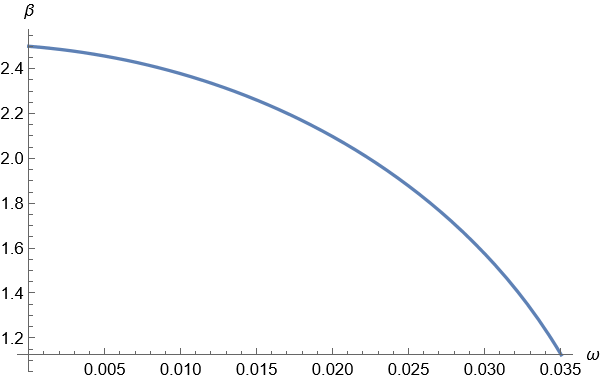}
	\caption{The relation between $\tilde{\beta}^{2}$ and $\omega$ when $s=2$, $\ell=2$, $m=1$ in quantum theory}
    \label{Fig:beta_s-2_n2_m1}
\end{figure}

In \cite{Brown:2024ajk}, the conformal dimension depends only on the quantum number $l$ for total orbital angular momentum. In the Kerr black hole, however, the angular dependence is via the spin-weighted spheroidal function whose eigenvalues depend on $a\omega$. Only for $a\omega=0$ the eigenvalue of spin-weighted spheroidal function is simply $\ell\left(\ell+1\right)-s(s+1)$. It is convenient to define the quantity $\beta=\sqrt{\frac{1}{4}-p_{1}+p_{4}}$, which is a function of $\omega$. In Figure \ref{Fig:beta_s-2_n2_m1}, we present the value of $\beta$ for  $s=2$, $\ell=2$, and $m=1$
For non-zero $\omega$, $\tilde{\beta} < l + \frac{1}{2}$, which implies that the effective conformal dimension is smaller than $1$. As a result, the product of the four $\Gamma$-functions in the correlators is larger than in the non-rotating case. As a result, for particles with positive $m$, that is, rotating in the same direction as the black hole, the radiation is enhanced. 
	
As expected, for the $m=0$ case the effect of rotation is negligible and the quantum radiation is suppressed  with respect to the semiclassical radiation -- mimicking the behaviour already reported in \cite{Brown:2024ajk}. For negative values of $m$ the expectation is that radiation will be suppressed. In fact, the upper bound of $\omega$ is very small while the conformal dimension cannot be very small. In some cases the upper bound might even be smaller than zero, so there will be no energy emission. 
	
\begin{figure}[t]
	\centering
	\includegraphics[width=0.5\textwidth]{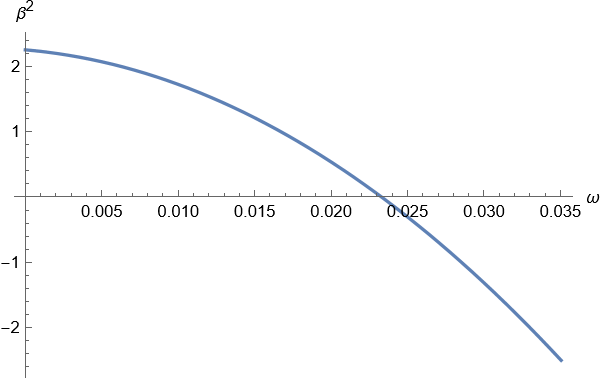}
	\caption{The relation between $\tilde{\beta}^{2}$ and $\omega$ for  $s=1$, $\ell=1$, $m=1$ in quantum-corrected theory}
    \label{Fig:s-1_l-1_m-1}
\end{figure}

Let us now consider $\beta$ for $s=1$, $\ell=1$, $m=1$, which is the dominant contribution to photon emission, and we plot in Figure \ref{Fig:s-1_l-1_m-1}. 
For large $\omega$, the value of $\tilde{\beta}^{2}$ becomes negative, so $\tilde{\beta}$ is imaginary, and $\frac{dE}{dtd\omega}$ will be a complex quantity. This breakdown of photon emission due to imaginary values of $\tilde{\beta}$ deserves further investigation elsewhere.    

\section{Conclusions}\label{Sec:Conclusions}

In this manuscript we have studied the effects that strong quantum fluctuations in the throat of near-extremal black holes have on the spectrum of Hawking radiation. We have found that in the quantum regime the emission spectrum is quite different from its semiclassical counterpart. This modification of Hawking radiation is naturally anticipated by the various computations of the effects that these quantum fluctuations have on the low-temperature thermodynamics of near-extremal black holes. Compared with the spherically symmetric case discussed in \cite{Brown:2024ajk}, we find two important cases. For particles without angular momentum, our results qualitatively agree with the main conclusion of \cite{Brown:2024ajk}, indicating a suppression of emission due to quantum effects in the throat. However, for particles carrying non-vanishing angular momentum in the same direction as the black hole, we found an enhancement with respect to the semiclassical result.

It is worth emphasizing that the instability we uncover is a new type of instability. We detect it effectively as a complex conformal dimension in the dual field theory interpretation. This instability is different from superradiance.

We leave a detailed description of the quantum-corrected evaporation of near-extremal rotating black holes within the Standard Model for a future work. There are a number of technical and conceptual obstacles, including the precise interpretation of some of the instabilities that show up in the form of complex conformal dimension for the dual operators. Phenomenologically, however, we anticipate that these corrections might have implications for bounds on the lifetime of primordial black holes.

Another interesting problem would be to consider the effect of quantum fluctuations for the evaporation of black holes in asymptotically AdS spacetimes which is quite different from the asymptotically flat case. It is still worth considering whether the AdS/CFT correspondence can provide a direct computational insight into these questions; some approaches to near-horizon rates were addressed in \cite{Nian:2020qsk} in the context of the Kerr/CFT correspondence; it would be interesting to connect that approach with the modifications studied here. 

Finally, it is important to arrive at a more systematic way to include quantum corrections in the whole spacetime, not just the near-horizon region. Recent advances in the computation of determinants in the full spacetime make us optimistic that it could be done \cite{Kapec:2023ruw, Kolanowski:2024zrq,Arnaudo:2024bbd}. 

\section*{Acknowledgments}
We  enjoy interesting conversations with Alfredo Gonz\'alez Lezcano, Matt Heydeman, Jim Liu, Xiao-Long Liu, Jun Nian, Cong-Yuan Yue, Jingchao Zhang. LPZ is partially supported by the U.S. Department of Energy under grant DE-SC0007859.

\appendix

\section{Eigenvalues of the spheroidal function} 
Eigenvalues of spin-weighted spheroidal function is difficult to compute. But thanks to the Black Hole Perturbation Toolkit \cite{BHPToolkit} it is possible for us to compute these complicated quantities.

To compute the eigenvalue in the equation
\begin{equation}
    \left( \frac{1}{\sin\theta}\partial_{\theta}\sin\theta \partial_{\theta}-\gamma^{2}\sin^{2}\theta-(\frac{m+s\cos\theta}{\sin\theta})^{2}-2\gamma s \cos\theta +s+2m\gamma+\ _{s}\lambda_{lm} \right)\ _{s}R_{lm}=0,
\end{equation}
we just need to use the Mathematica code $SpinWeightedSpheroidalEigenvalue[s, l, m, \gamma]$ to compute $\ _{s}\lambda_{lm}$.

We consider an example where $s=0$,$n=m=1$, and we use $a\omega=0.5$ as the result in classical gravity, so $\tilde{\beta}=0.670606$ so we know that the difference between $\tilde{\beta}$ and $l+\frac{1}{2}$ cannot be ignored.

\section{Confluent Heun's Equation} \label{appendix:CHE}

In this appendix we summarize some of the remarkable results obtained in \cite{Bonelli:2021uvf} (see also \cite{Arnaudo:2024rhv}), including a slight modification to adjust them to our more general situation. Mathematically the question is how to relate the coefficients of expansions of the solution to the wave equation around specific regions (near-horizon and asymptotic). This problem appears in various areas of physics and it has recently been noted that one can translate the advanced in the context of conformal field theories into more generic situations. 

Let us first show that, after separation of variables, the radial equation is equivalent to a confluent Heun's equation. To verify this, we define a new function $\phi(r)$ and subsequently a new radial variable $z$ as 
    \begin{equation}
    	\phi(r)=\Delta^{\frac{s+1}{2}}R(r), \qquad 
    	z=\frac{r-r_{-}}{r_{+}-r_{-}}.
    \end{equation}
    
The original equation for $\phi(r)$ is  
    \begin{equation}
    	\Delta \partial_{r}^{2}\phi+((1-s^{2})\frac{(r-M)^{2}}{\Delta}-s-1+\frac{K_{eff}^{2}-2is(r-M)K_{eff}}{\Delta}-\Lambda+4is\omega r) R=0,
    \end{equation}
and, in the new coordinate, $z$, becomes
    \begin{equation}
    	\partial_{z}^{2}\phi+\frac{P(z)}{z^{2}(z-1)^{2}}\phi=0,
    \end{equation}
where $P(z)$ is a degree-4 polynomial of $z$ 
    \begin{equation}
    	\begin{split}
    	P(z)=&(1-s^{2})(z-\frac{1}{2})^{2}
    	\\&+\frac{(\omega (r_{-}+(r_{+}-r_{-})z)^{2}+\omega a^{2}-ma-eQ(r_{-}+(r_{+}-r_{-})z))^{2}}{(r_{+}-r_{-})^{2}}
    	\\&-2is\frac{(z-\frac{1}{2})(\omega (r_{-}+(r_{+}-r_{-})z)^{2}+\omega a^{2}-ma-eQ(r_{-}+(r_{+}-r_{-})z))}{r_{+}-r_{-}}
    	\\&+(r_{+}-r_{-})^{2}z(z-1)(-s-1-\Lambda+4is\omega(r_{-}+(r_{+}-r_{-})z)).
    	\end{split}
    \end{equation}
Therefore, the equation satisfied by  $\phi(z)$ is a confluent Heun's equation. It is sometimes convenient to express $P(z)$ explicitly as 
	\begin{equation}
		P(z)=\sum_{i=0}^{4}A_{i}z^{i},
	\end{equation}
the parameters $\{A_{i}\}$ will play an important role. 

\subsection{CHE is equivalent to BPZ equation}

Recent progress on the confluent Heun's equation was achieved in the context of certain formal properties of supersymmetric field theories. The authors of \cite{Bonelli:2021uvf} were able to translate these developments into gravitational aspects. 
 
Now we focus on chiral conformal field theory. In this theory, we work with conformal momenta related to conformal weight by $\Delta=\frac{Q^{2}}{4}-\alpha^{2}$. The weight of degenerate Verma modules is $\Delta_{r,s}=\frac{Q^{2}}{4}-\alpha_{r,s}^{2}$. Here $\alpha_{r,s}=-\frac{br}{2}-\frac{s}{2b}$, and $Q=b+\frac{1}{b}$, and the central charge $c=1+6Q^{2}$. For $r=2$, $s=1$, $\Delta_{2,1}=-\frac{1}{2}-\frac{3}{4}b^{2}$, the EoM of $\Psi_{2,1}$ is
	\begin{equation}
		(\frac{1}{b^{2}}L_{-1}^{2}+L_{-2})\Phi_{2,1}(z)=0
	\end{equation}
	
	Now we define the conformal block
	\begin{equation}
		\Psi(z)=<\Delta,\Lambda_{0},m_{0}|\Psi_{2,1}(z)V_{2}(1)|\Delta_{1}>
	\end{equation}
	
	Here $V_{2}(1)$ is a primary operator of weight $\Delta_{2}=\frac{Q^{2}}{4}-\alpha_{2}^{2}$ inserted at $z=1$. $|\Delta_{1}>$ is a primary state of weight $\Delta_{1}=\frac{Q^{2}}{4}-\alpha_{1}^{2}$.
	
	The EoM of $\Psi(z)$ is
	\begin{equation}
		<\Delta,\Lambda_{0},m_{0}|(\frac{1}{b^{2}}\partial_{z}^{2}+L_{-2})\Psi_{2,1}(z)V_{2}(1)|V_{1}>=0
	\end{equation}
	
	we simplify this equation and get the BPZ equation
	\begin{equation}
		(\frac{1}{b^{2}}\partial_{z}^{2}-\frac{1}{z}\partial_{z}-\frac{1}{z(z-1)}(z\partial_{z}-\Lambda_{0}\partial_{\Lambda_{0}}+\Delta_{2,1}+\Delta_{2}+\Delta_{1}-\Delta)+\frac{\Delta_{2}}{(z-1)^{2}}+\frac{\Delta_{1}}{z^{2}}+\frac{m_{0}\Lambda_{0}}{z}+\Lambda_{0}^{2})\Psi(z)=0
	\end{equation}
	
	Now we want to simplify this equation using double-scaling limit. We define $\epsilon_{2}=\hbar b\rightarrow 0$. Meanwhile, we fix $\epsilon_{1}=\frac{\hbar}{b}$,$\hat{\Delta}=\hbar^{2}\Delta$,  $\hat{\Delta}_{1}=\hbar^{2}\Delta_{1}$, $\hat{\Delta}_{2}=\hbar^{2}\Delta_{2}$, $\Lambda=2i\hbar\Lambda_{0}$, $m_{3}=\frac{i}{2}\hbar m_{0}$
	
	the z dependence of $\Psi_{z}$ is only in subleading order, $\Psi(z)$ has the form 
	\begin{equation}
		\Psi(z)\propto e^{\frac{1}{\epsilon_{1}\epsilon_{2}}(\mathcal{F}^{inst}(\epsilon_{1})+\epsilon_{2}\mathcal{W}(z;\epsilon_{1})+\mathcal{O}(\epsilon_{2}^{2}))}
	\end{equation}
	
	so the EoM of $\Psi(z)$ is
 	\begin{equation}
		(\epsilon_{1}^{2}\partial_{z}^{2}-\frac{1}{z(z-1)}(-\Lambda \partial_{\Lambda}\mathcal{F}^{inst}+\hat{\Delta}_{2}+\hat{\Delta}_{1}-\hat{\Delta})+\frac{\hat{\Delta}_{2}}{(z-1)^{2}}+\frac{\hat{\Delta}_{1}}{z^{2}}-\frac{m_{3}\Lambda}{z}-\frac{\Lambda^{2}}{4})\Psi(z)=0
	\end{equation}
	
	This is a CHE. Now we define $\hat{\Delta}_{i}=\frac{1}{4}-a_{i}^{2}$ , $E=a^{2}-\Lambda \partial_{\Lambda}\mathcal{F}^{inst}$In addition, we set $\epsilon_{1}=1$, so we finally get the expression for each $A_{i}$
	\begin{equation}
		\begin{split}
			&A_{0}=\frac{1}{4}-a_{1}^{2}
			\\&A_{1}=-\frac{1}{4}+E+a_{1}^{2}-a_{2}^{2}-m_{3}\Lambda
			\\&A_{2}=\frac{1}{4}-E+2m_{3}\Lambda-\frac{\Lambda^{2}}{4}
			\\&A_{3}=-m_{3}\Lambda+\frac{\Lambda^{2}}{2}
			\\&A_{4}=-\frac{\Lambda^{2}}{4}
		\end{split}
	\end{equation}
	
	Now we know $\{A_{i}\}$ so we can get each parameters in BPZ equation.
	\subsection{Solutions of BPZ equation}
	We solve the BPZ equation near $z=1$ and get the t channel solution
	\begin{equation}
		f_{\alpha_{2+}}^{(t)}=<\Delta_{\alpha},\Lambda_{0},m_{0}|V_{\alpha_{2+}}(1)|\Delta_{\alpha_{1}}>(z-1)^{\frac{bQ+2b\alpha_{2}}{2}}(1+\mathcal{O}(z-1))
	\end{equation}
	\begin{equation}
		f_{\alpha_{2-}}^{(t)}=<\Delta_{\alpha},\Lambda_{0},m_{0}|V_{\alpha_{2-}}(1)|\Delta_{\alpha_{1}}>(z-1)^{\frac{bQ-2b\alpha_{2}}{2}}(1+\mathcal{O}(z-1))(1+\mathcal{O}(z-1))
	\end{equation}
	
	and we solve the BPZ equation in $z>>1$ and get the u channel solution
	\begin{equation}
		f_{\alpha+}^{(u)}=\Sigma_{\pm}<\Delta_{\alpha_{+}},\Lambda_{0},m_{0\pm}|V_{\alpha_{2}}(1)|\Delta_{\alpha_{1}}>\mathcal{A}_{\alpha_{+},m_{0\pm}}e^{\pm \frac{\Lambda z}{2}})(\pm \Lambda)^{-\frac{1}{2}\pm m_{3}+b\alpha_{+}}z^{\frac{1}{2}(bQ-1\pm 2m_{3})}(1+\mathcal{O}(z^{-1}))
	\end{equation}
	\begin{equation}
		f_{\alpha-}^{(u)}=\Sigma_{\pm}<\Delta_{\alpha_{-}},\Lambda_{0},m_{0\pm}|V_{\alpha_{2}}(1)|\Delta_{\alpha_{1}}>\mathcal{A}_{\alpha_{-},m_{0\pm}}e^{\pm \frac{\Lambda z}{2}})(\pm \Lambda)^{-\frac{1}{2}\pm m_{3}-b\alpha_{-}}z^{\frac{1}{2}(bQ-1\pm 2m_{3})}(1+\mathcal{O}(z^{-1}))
	\end{equation}
	
	We can connect these solutions by
	\begin{equation}
		f_{i}^{(t)}(z)=M_{ij}f_{j}^{(u)}(z)
	\end{equation}
	
	Here $i=\alpha_{2\pm}$ and $j=\alpha_{\pm}$
	
	According to DOZZ formula, the matrix elements are
	\begin{equation}
		\begin{split}
			&M_{\alpha_{2+},\alpha_{+}}=\frac{ \Gamma(-2b\alpha)\Gamma(1+2b\alpha_{2})  }{ \Gamma(\frac{1}{2}+b(\alpha_{1}+\alpha_{2}-\alpha))\Gamma(\frac{1}{2}+b(-\alpha_{1}+\alpha_{2}-\alpha))  }
			\\&M_{\alpha_{2-},\alpha_{-}}=\frac{ \Gamma(2b\alpha)\Gamma(1-2b\alpha_{2})  }{ \Gamma(\frac{1}{2}+b(\alpha_{1}-\alpha_{2}+\alpha))\Gamma(\frac{1}{2}+b(-\alpha_{1}-\alpha_{2}+\alpha))  }
			\\&M_{\alpha_{2+},\alpha_{-}}=\frac{ \Gamma(2b\alpha)\Gamma(1+2b\alpha_{2})  }{ \Gamma(\frac{1}{2}+b(\alpha_{1}+\alpha_{2}+\alpha))\Gamma(\frac{1}{2}+b(-\alpha_{1}+\alpha_{2}+\alpha))  }
			\\&M_{\alpha_{2-},\alpha_{+}}=\frac{ \Gamma(-2b\alpha)\Gamma(1-2b\alpha_{2})  }{ \Gamma(\frac{1}{2}+b(\alpha_{1}-\alpha_{2}-\alpha))\Gamma(\frac{1}{2}+b(-\alpha_{1}-\alpha_{2}-\alpha))  }
		\end{split}
	\end{equation}

    to solve the connection problem in Teukolsky's equation, we assume $a_{i}=\hbar \alpha_{i}=b\alpha_{i}$ for $\epsilon_{1}=\frac{\hbar}{b}=1$, so we write down the matrix

    \begin{equation}\label{Eq:DOZZ}
		\begin{split}
			&M_{a_{2+},a_{+}}=\frac{ \Gamma(-2a)\Gamma(1+2a_{2})  }{ \Gamma(\frac{1}{2}+a_{1}+a_{2}-a)\Gamma(\frac{1}{2}-a_{1}+a_{2}-a)  }
			\\&M_{a_{2-},a_{-}}=\frac{ \Gamma(2a)\Gamma(1-2a_{2})  }{ \Gamma(\frac{1}{2}+a_{1}-a_{2}+a)\Gamma(\frac{1}{2}-a_{1}-a_{2}+a)  }
			\\&M_{a_{2+},a_{-}}=\frac{ \Gamma(2a)\Gamma(1+2a_{2})  }{ \Gamma(\frac{1}{2}+a_{1}+a_{2}+a)\Gamma(\frac{1}{2}-a_{1}+a_{2}+a)  }
			\\&M_{a_{2-},a_{+}}=\frac{ \Gamma(-2a)\Gamma(1-2a_{2})  }{ \Gamma(\frac{1}{2}+a_{1}-a_{2}-a)\Gamma(\frac{1}{2}-a_{1}-a_{2}-a)  }
		\end{split}
	\end{equation}
	
	Now we use the boundary condition in the Killing horizon, there is only ingoing wave near the horizon, so the near horizon solution is
	\begin{equation}
		\psi(z)=(z-1)^{\frac{1}{2}+a_{2}}
	\end{equation}
	
	and connect this solution with solution in far horizon region
	\begin{equation}
		\begin{split}
		\psi(z)&=\frac{1}{<\Delta_{\alpha},\Lambda_{0},m_{0}|V_{\alpha_{2+}}(1)|\Delta_{\alpha_{1}}>}(M_{\alpha_{2+},\alpha_{-}}f_{\alpha_{-}}^{(u)}(z)+M_{\alpha_{2+},\alpha_{+}}f_{\alpha_{+}}^{(u)}(z))
		\\&=M_{\alpha_{2+},\alpha_{-}}\Lambda^{-\frac{1}{2}-a}\Sigma_{\pm}\mathcal{A}_{\alpha_{-},m_{3\pm}}e^{\pm \frac{\Lambda z}{2}}(\Lambda z)^{\pm m_{3}}\frac{ <\Delta_{\alpha_{-}},\Lambda_{0},m_{0 +}|V_{\alpha_{2}}(1)|\Delta_{\alpha_{1}}>  }{  <\Delta_{\alpha},\Lambda_{0},m_{0 }|V_{\alpha_{2+}}(1)|\Delta_{\alpha_{1}}>  }(1+\mathcal{O}(z^{-1}))+(\alpha \rightarrow -\alpha)
		\end{split}
	\end{equation}

\section{Semiclassical limit of the quantum-corrected emission of gravitons}

Our goal in this appendix is to show the necessary steps demonstrating that the quantum-corrected emission rate for gravitons reduces to the semi-classical one in the appropriate limit. In subsection \ref{SubSec:SemiclassicalLimit-Scalar} we presented the case for emission of scalar particles; the details for photon emission are similar to those presented here. Agreement with the semi-classical expression is an important sanity check for the quantum-corrected expressions and allows us to fix some ambiguities in normalizations. 

With the information provided in the main text, we can write the quantum-corrected emission rate for gravitons ($s=2$) as follows, 
	\begin{equation}\label{Eq:GravitonEmission-total}
		\begin{split}
			\frac{dE}{dtd\omega}&=A_{s}\frac{8\omega^{5}(r_{+}-r_{-})^{4-\sqrt{1-4p_{1}+4p_{4}}}}{\pi}L_{2}^{2\sqrt{1-4p_{1}+4p_{4}}}\times
			\\&2\times \left|\frac{ \Gamma(\frac{1}{2}+m_{3}+\sqrt{E})\Gamma(\frac{1}{2}+a_{1}+a_{2}+\sqrt{E})\Gamma(\frac{1}{2}-a_{1}+a_{2}+\sqrt{E})  }{  \Gamma(1+2\sqrt{E})\Gamma(2\sqrt{E})\Gamma(1+2a_{2}))(-2i\omega(r_{+}-r_{-})^{-\sqrt{E}+m_{3} }}\right|^{2}\times
			\\&\left|\frac{\Gamma(\sqrt{1-4p_{1}+4p_{4}})\Gamma(1-2i\sqrt{p_{3}})}{\Gamma(\frac{1}{2}+\frac{1}{2}\sqrt{1-4p_{1}+4p_{4}}+i\sqrt{p_{3}-p_{2}+p_{1}}-i\sqrt{p_{3}})\Gamma(\frac{1}{2}--i\sqrt{p_{3}-p_{2}+p_{1}}-i\sqrt{p_{3}}+\frac{1}{2}\sqrt{1-4p_{1}+4p_{4}}}\right|^{2}\times
			\\&\frac{C}{\pi^{2}} \sinh(2\pi\sqrt{2C(E_{2}-\frac{j^{2}n_{2}^{2}}{2K_{r}}-\frac{e^{2}}{2K_{e}})}) \times
			\\&\frac{ \Gamma (\frac{1}{2}+\frac{1}{2}\sqrt{1-4p_{1}+4p_{4}} \pm i(\sqrt{2CE_{1}} \pm \sqrt{2C(E_{2}-\frac{j^{2}n_{2}^{2}}{2K}-\frac{e^{2}}{2K_{e}})}))}{(2C)^{1+\sqrt{1-4p_{1}+4p_{4}}}\Gamma(1+\sqrt{1-4p_{1}+4p_{4}})},
		\end{split}
	\end{equation}
	and
	\begin{equation}
		A_{2}=\frac{2^{4}\Gamma(6)}{\Gamma^{2}(5)}.
	\end{equation}
Let us first note that some of the  $\Gamma$ functions in the second and third lines cancel each other. Indeed, one can verify that 
	\begin{equation}
		\sqrt{1-4p_{1}+4p_{4}}=2\sqrt{E},
	\end{equation}
which implies that 	$\Gamma(2\sqrt{E})$ in the denominator of the second line cancels with $\Gamma(\sqrt{1-4p_{1}+4p_{4}})$ in the numerator of the third line. Similarly, 	$\Gamma(1+2a_{2})$ in the second line cancels with $\Gamma(1-2i\sqrt{p_{3}})$ in the third line. We also use the approximation $\sqrt{p_{3}-p_{2}+p_{1}}-\sqrt{p_{3}}\approx is+(eQ-2r_{+}\omega)$ and consider the branch  $\sqrt{p_{3}}=\frac{\omega_{\rm eff}}{2\pi T}-i\frac{s}{2}$. As a result,
	\begin{equation}
		\sqrt{p_{3}-p_{2}+p_{1}}\approx \frac{\omega_{\rm eff}}{2\pi T}+i\frac{s}{2} -i(eQ-2r_{+}\omega).
	\end{equation}
	
Analogously, $\Gamma(\frac{1}{2}+\frac{1}{2}\sqrt{1-4p_{1}+4p_{4}}+i\sqrt{p_{3}-p_{2}+p_{1}}-i\sqrt{p_{3}})$ in the third line cancels with $\Gamma(\frac{1}{2}+a_{1}+a_{2}+\sqrt{E})$ in the second line, and they both equal 
	\begin{equation}
		\Gamma(\frac{1}{2}+\sqrt{E}-s-2iM\omega).
	\end{equation}
	
Finally, $\Gamma(\frac{1}{2}+\frac{1}{2}\sqrt{1-4p_{1}+4p_{4}}-i\sqrt{p_{3}-p_{2}+p_{1}}-i\sqrt{p_{3}})$ in the third line cancels with $\Gamma(\frac{1}{2}-a_{1}+a_{2}+\sqrt{E})$ in the second line, and they both equal to 
	\begin{equation}
		\Gamma(\frac{1}{2}+\sqrt{E}-i\frac{\omega_{\rm eff}}{2\pi T}+2iM\omega).
	\end{equation}

To find the normalization, we recall that assuming $a=M=r_{+}$ in the very low temperature limit and using the notation of \cite{Bonelli:2021uvf}
	\begin{equation}
		\begin{split}
			E=&\frac{1}{4}+\lambda+s(s+1) +a^{2}\omega^{2}-8a^{2}M^{2}
			\\&\approx \frac{1}{4}+\lambda+s(s+1)-7a^{2}\omega^{2},
		\end{split}
	\end{equation}
leading to $1-4p_{1}+4p_{4}\approx 4E$. The asymptotic value of the radial function is thus 	
	\begin{equation}
		\begin{split}
			Y_{s} \xrightarrow{x \to \infty}& N\, A_{1} \left\{ \right.
			\\&\frac{\Gamma(\sqrt{1-4p_{4}+4p_{3}})\Gamma(1-2i\sqrt{p_{2}})}{\Gamma(\frac{1}{2}+\frac{1}{2}\sqrt{1-4p_{4}+4p_{3}}-i\sqrt{p_{2}}+i\sqrt{p_{4}-p_{1}+p_{2}})\Gamma(\frac{1}{2}-i\sqrt{p_{2}}-i\sqrt{p_{4}-p_{1}+p_{2}}-\frac{1}{2}\sqrt{1-4p_{4}+4p_{3}})}\times
			\\&x^{-\frac{1}{2} - \frac{1}{2}\sqrt{1 + 4 p_{3} - 4 p_{4}} } +
			\\&\frac{\Gamma(-\sqrt{1-4p_{4}+4p_{3}})\Gamma(1-2i\sqrt{p_{2}})}{\Gamma(\frac{1}{2}-\frac{1}{2}\sqrt{1-4p_{4}+4p_{3}}-i\sqrt{p_{2}}+i\sqrt{p_{4}-p_{1}+p_{2}})\Gamma(\frac{1}{2}-i\sqrt{p_{2}}+i\sqrt{p_{4}-p_{1}+p_{2}}-\frac{1}{2}\sqrt{1-4p_{4}+4p_{3}})}\times
			\\& \left. x^{-\frac{1}{2} + \frac{1}{2}\sqrt{1 + 4 p_{3} - 4 p_{4}} }  \right\}.
		\end{split}
	\end{equation}

The simplification of the correlator (product of four $\Gamma$ functions) is by now standard and we will not repeat it here, except for noting that  it requires a few approximations:  $M\omega<<1$, $E_{i}>>E_{\rm brk}$, so $E_{i}C>>1$. Since for the computations of the total emission we need $\int d\omega \,\frac{dE}{dtd\omega}$, the dominant contribution comes from  $\omega \propto\sqrt{E_{i}C}$ and, therefore, we further use the approximate condition $E_{i}C\propto \omega^{2}>>1$.
	
Since the second line in the general formula (\ref{Eq:GravitonEmission-total}) appears in the semi-classical formula, we focus on the third line. In the regime with  $M\omega<<1$, $\sqrt{E}=n+\frac{1}{2}$ the third line becomes
	\begin{equation}
		\begin{split}
			&\left|   \frac{\Gamma(2n+1)\Gamma(1-s-i\frac{\omega_{\rm eff}}{2\pi T_{H}}) }{\Gamma(1+l-i\frac{\omega_{\rm eff}}{2\pi T_{H}})\Gamma(1+l-s) }   \right|^{2}
			\\=&\left|   \frac{\Gamma(-1-i\frac{\omega_{\rm eff}}{2\pi T_{H}}) }{\Gamma(3-i\frac{\omega_{\rm eff}}{2\pi T_{H}}) }   \right|^{2}\Gamma^{2}(5)
			\\=&\left|   \frac{1 }{  (2-i\frac{\omega_{\rm eff}}{2\pi T_{H}})(1-i\frac{\omega_{\rm eff}}{2\pi T_{H}})(-i\frac{\omega_{\rm eff}}{2\pi T_{H}})          (-1-i\frac{\omega_{\rm eff}}{2\pi T_{H}})}   \right|^{2}\Gamma^{2}(5)
			\\=&\frac{1}{(4+(\frac{\omega_{\rm eff}}{2\pi T_{H}})^{2}) (1+(\frac{\omega_{\rm eff}}{2\pi T_{H}})^{2})^{2}   (\frac{\omega_{\rm eff}}{2\pi T_{H}})^{2} }\Gamma^{2}(5)
		\end{split}
	\end{equation}
	in the first line,  we assume $\sqrt{E}=n+\frac{1}{2}$ and ignore terms of the form $2iM\omega$.
    
Let us denote by $6\Gamma$ the following factor
	\begin{equation}
	6\Gamma \equiv \left|\frac{ \Gamma(\frac{1}{2}+m_{3}+\sqrt{E})\Gamma(\frac{1}{2}+a_{1}+a_{2}+\sqrt{E})\Gamma(\frac{1}{2}-a_{1}+a_{2}+\sqrt{E})  }{  \Gamma(1+2\sqrt{E})\Gamma(2\sqrt{E})\Gamma(1+2a_{2}))(-2i\omega(r_{+}-r_{-})^{-\sqrt{E}+m_{3} }}\right|^{2}
	\end{equation}
	
The quantum-corrected emission rate can now be written as 
\begin{equation}
		\begin{split}
			\frac{dE}{dtd\omega}=&A_{2}\frac{8\omega^{5}(r_{+}-r_{-})^{-1}}{\pi}L_{2}^{10}\times2\times 6\Gamma \times \frac{1}{(4+(\frac{\omega_{\rm eff}}{2\pi T_{H}})^{2}) (1+(\frac{\omega_{\rm eff}}{2\pi T_{H}})^{2})^{2}   (\frac{\omega_{\rm eff}}{2\pi T_{H}})^{2} }\Gamma^{2}(5)\times
			\\&\frac{1}{\Gamma(6)}\omega_{\rm eff}(\omega_{\rm eff}^{2}+4\pi^{2}T_{H})(\omega_{\rm eff}^{2}+16\pi^{2}T_{H}^{2})\times \frac{1}{e^{  \frac{\omega_{\rm eff}}{T_{H}}   }-1} 
			\\=&A_{2}\frac{16\omega^{2}(2M^{2})^{5}}{\pi (r_{+}-r_{-})}\times 6\Gamma \times \frac{(2\pi T_{H})^{6}}{\omega_{\rm eff}(1+(\frac{\omega_{\rm eff}}{2\pi T_{H}})^{2})} \times \frac{1}{e^{  \frac{\omega_{\rm eff}}{T_{H}}   }-1}
		\end{split}
	\end{equation}
where  $L_{2}^{2}=2M^{2}$ was used the second line. We are now ready to compare this result with the expression for the semiclassical result. Taking $A_{s=2}=\frac{2^{4}\Gamma(6)}{\Gamma^{2}(5)}$, leads precisely to the semiclassical expression: 
	\begin{equation}
		\left.\frac{dE}{dtd\omega}\right|_{\rm 
        Quantum-corrected}\to \frac{1}{2\pi} \sigma \times \frac{1}{e^{  \frac{\omega_{\rm eff}}{T_{H}}   }-1},
	\end{equation}
    where $\sigma$ is the greybody factor as reported in \cite{Teukolsky:1973ha,Teukolsky:1974yv}.
\bibliographystyle{JHEP}
\bibliography{Hawk-Kerr}

\end{document}